%% file: toplevel.tex
\documentclass[accepted]{article}

\usepackage{tikz}
\usepackage{xcolor}
\usepackage[ruled,vlined]{algorithm2e}
\usepackage{microtype}
\usepackage{graphicx}
\usepackage{subcaption}
\usepackage{booktabs} 
\usepackage{pifont}

\usepackage[colorlinks=true, urlcolor=blue]{hyperref}
\usepackage[T1]{fontenc}
\usepackage{mathtools}
\usepackage{pifont}
\usepackage{array,booktabs,ragged2e}
\usepackage{enumitem}
\usepackage{listings}
\usepackage{xcolor}
\usepackage{float}
\usepackage{tablefootnote}
\usepackage{enumitem}
\usepackage{listings}
\usepackage{amsmath}
\usepackage{amssymb}
\usepackage[sans]{dsfont}
\usepackage{titlesec}
\newcommand{\app}[1]{{\tt#1}}
\newcommand{\model}[1]{{\tt#1}}
\newcommand{\machine}[1]{{\tt#1}}

\newcolumntype{R}[1]{>{\RaggedLeft\arraybackslash}p{#1}}

\newcommand{\nvidia}[0]{NVIDIA} 

\newcommand{\tickmark}[0]{\ding{52}}
\newcommand{\xmark}[0]{\ding{53}}
\newcommand{\xlm}[0]{xLM}

\newcommand{\plangpuonly}[0]{{\em GPU-only}}
\newcommand{\planstatic}[0]{{\em Static}}
\newcommand{\plandynamic}[0]{{\em Dynamic}}
\newcommand{\ukv}[0]{{\em ukv}}
\newcommand{\nukv}[0]{{\em nukv}}
\newcommand{\batchavg}[0]{2.3$\times$}
\newcommand{\batchmax}[0]{8.2$\times$}

\newcommand*\circled[1]{%
    \tikz[baseline=(char.base)]{%
        \node[shape=circle, fill, inner sep=2pt] (char) {\textcolor{white}{#1}};%
    }%
}

\usepackage{eso-pic}
\usepackage[accepted]{mlsys2026}

\mlsystitlerunning{Efficient, VRAM-Constrained \xlm\ Inference On Clients}

\begin{document}

\AddToShipoutPictureBG*{ 
  \AtPageUpperLeft{
    \hspace*{\paperwidth-65pt\relax}
    \raisebox{-68pt}{
      \llap{
      \href{https://www.acm.org/publications/policies/artifact-review-and-badging-current}{\includegraphics[height=65pt]{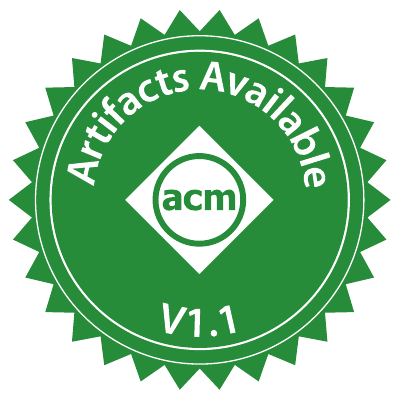}}
\hspace{1pt}
      \href{https://www.acm.org/publications/policies/artifact-review-and-badging-current}{\includegraphics[height=65pt]{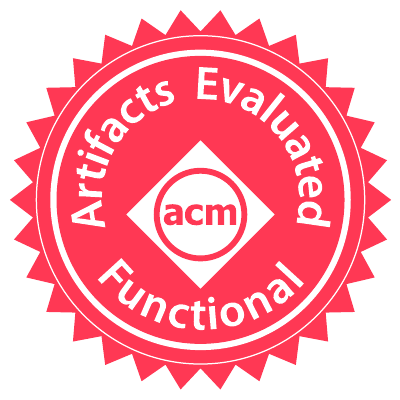}}
\hspace{1pt}
      \href{https://www.acm.org/publications/policies/artifact-review-and-badging-current}{\includegraphics[height=65pt]{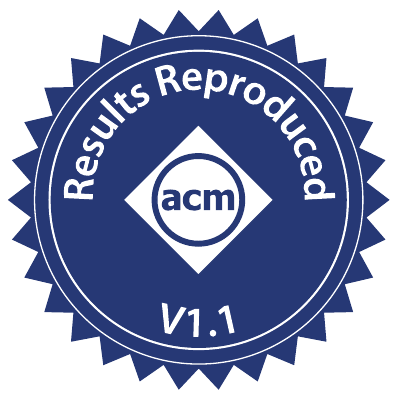}}
     }
    }
  }
}

\twocolumn[
\mlsystitle{Efficient, VRAM-Constrained \xlm\ Inference On Clients}



\mlsyssetsymbol{equal}{*}

\begin{mlsysauthorlist}
\mlsysauthor{Aditya Ukarande}{ed}
\mlsysauthor{Deep Shekhar}{to}
\mlsysauthor{Marc Blackstein}{ed}
\mlsysauthor{Ram Rangan}{to}
\end{mlsysauthorlist}

\mlsysaffiliation{ed}{NVIDIA Corporation, Santa Clara, CA, USA}
\mlsysaffiliation{to}{NVIDIA Graphics Pvt Ltd., Bangalore, India}

\mlsyscorrespondingauthor{Aditya Ukarande, Ram Rangan}{\{aukarande, rrangan\}@nvidia.com}

\mlsyskeywords{local inference, on-device inference, inference on consumer GPUs, pipelined sharding, benchmark-profile-guided inference scheduling, CPU-GPU hybrid scheduling, sub-layer model sharding, TTFT and throughput optimization, accuracy-preserving, lossless}

\vskip 0.3in

\begin{abstract}

\include{abstract}

\end{abstract}
]



\printAffiliationsAndNotice{}  

\input{intro}
\input{background}
\input{related}

\input{pipeshard}

\input{vlmopt}

\input{methodology}

\input{eval}

\input{discuss}

\input{conc}

\section*{Acknowledgments}
Many thanks to our anonymous reviewers, Gaurav Garg, and Vivek Agrawal for their excellent feedback that has improved the quality of the paper. 
We thank our artifact evaluators for their constructive feedback and collaborative efforts in improving our artifact.
Thanks to Vidya Murali, Andrew Edelsten, Emmett Kilgariff, John Spitzer and the entire Applied Architecture team for their support.

\bibliography{paper}
\bibliographystyle{mlsys2026}

\appendix
\include{artifactappendix}

\end{document}

%% file: abstract.tex
To usher in the next round of client AI innovation, there is an urgent need to enable efficient, lossless inference of high-accuracy large language models~(LLMs) and vision language models~(VLMs), jointly referred to as {\xlm}s, on client systems. 
This means efficient support for: a)  interactive as well as batch modes, b) high-resolution VLM inference, c) dense and mixture-of-experts~(MoE) LLMs, and d) adapting to system conditions (CPU thread count, CPU-GPU interconnect bandwidth, and video memory~(VRAM) budget) and inference conditions (phase of execution and context size). While recent CPU-GPU hybrid scheduling techniques show promise, to the best of our knowledge, no single product handles all of the above. 

In this paper, we address this problem with \emph{pipelined sharding}, a novel, benchmark-profile-guided CPU-GPU hybrid scheduling technique to achieve efficient, VRAM-constrained inference for both dense and mixture-of-experts~(MoE) LLMs. 
Using a combination of model sharding at the sub-layer level, CPU offloading, pipelined copy-compute, and prioritized tensor placement in VRAM, it optimizes both time-to-first-token~(TTFT) and tokens per second~(TPS) metrics, while flexibly adapting to system and inference conditions. 
For efficient, high-accuracy VLM inference, we combine pipelined sharding with a llama.cpp implementation of three well-understood prior ideas~(jointly called \emph{VLMOpt}), namely, vision tensor CPU offloading, flash attention, and  vision and language model VRAM overlap avoidance. 

These enhancements are targeted at improving client \xlm\ inference in future releases of two important \nvidia\ products - the In-Game Inferencing software development kit~(IGI SDK) and the Cosmos-Reason1~(CR1) physical AI reasoning VLM.  
Highlights from our rigorous evaluation spanning multiple models and client systems include:  for interactive use, TTFT improves by up to 6.7$\times$ and TPS by up to 30$\times$ for LLMs, and CR1 inference's VRAM demand is down by 10$\times$, while in batched mode, throughput improves by up to \batchmax{},  all compared to their respective aggressive baselines.

%% file: intro.tex
\section{Introduction}
\label{sec:intro}

Today, autoregressive transformer-based large language models~(LLMs) and their close cousins, vision language models~(VLMs), underpin several modern AI applications, from AI assistants - gaming~\cite{gamecopilot}, home~\cite{ha-llmvision}, delivery robots~\cite{ottonomy} -  to various intelligent gaming initiatives~\cite{nvace}.  Even though popular cloud services offer high-accuracy LLMs and VLMs (jointly referred to as  {\xlm}s in this paper) at impressive throughputs, high cloud costs and privacy concerns~\cite{kshetri:23:itprofessional, iqbal:24:platformsecurity} encourage developers and users to look for client-based alternatives.  

In recent years, the open-source ecosystem~\cite{paszke:19:neurips, llamacpp, kwon:23:sosp} and the research community have contributed greatly towards enabling \xlm\ inference on client systems. 
However, the market still lacks products for efficient inference of high-accuracy {\xlm}s at flexible, user-specified video memory~(VRAM) budgets on client systems. Through this work, we aim to address this limitation and foster the next round of client AI innovation by: a) enabling game developers to integrate efficient, high-accuracy {\xlm} inference at any desired VRAM budget via the NVIDIA In-Game Inferencing~(IGI) SDK, and b) enabling physical AI developers to run efficient, high-resolution inference on NVIDIA's Cosmos-Reason1~(CR1)~\cite{nvcosmos} VLM on VRAM-constrained client systems.  For LLMs, high accuracy means running high parameter count models, both dense and sparse mixture-of-experts~(MoE) ones. For VLMs, high accuracy means running inference on high-resolution images to detect small/distant features and text along with high parameter count language decoders~\cite{vasu:25:cvpr, bai:25:tr}.  In order for these models to be practically useful, their time-to-first-token~(TTFT) must be low enough and tokens per second~(TPS) must be high enough to meet user needs.  

The single biggest challenge for efficient inference of  {\xlm}s on client systems is the limited VRAM availability in client GPUs.   A variant of this problem is to target efficient, high-accuracy \xlm\ inference at \emph{any} user-specified VRAM budget, and achieve the best TTFT and TPS for a given model at that specified VRAM budget. For NVIDIA, this is important for two reasons: 1) users looking to make the most of their client systems by running \xlm\ inference concurrently alongside other GPU applications (e.g., creative applications, video games), and 2) application developers, in order to present a uniform experience to their end users, could require the same quality \xlm\ to be runnable across a spectrum of GPUs with varying VRAM availability~(e.g. game assistant developers).

\begin{table*}[h]
\vspace{-0.05in}
\scriptsize
\centering
\begin{tabular}{|c|p{0.4in}|p{0.55in}|p{0.4in}|p{0.4in}|p{0.6in}|p{0.5in}|p{0.37in}|p{0.53in}|}
\hline
Technique & Optimizes TTFT & Works Well for Batch Size=1& Optimizes dense LLMs & Optimizes MoE LLMs & Different Priorities for Attention and FFN & KV Cache Sharding  & Optimizes VLMs &  Evaluated on Client Systems  \\ 
\hline
\hline
TwinPilots~\cite{yu:24:systor} & \xmark & \xmark & \tickmark & \xmark & \xmark & \xmark & \xmark & \xmark \\
\hline
HeteGen~\cite{zhao:24:mlsys} & \xmark & \tickmark & \tickmark & \xmark & \xmark & \xmark & \xmark &  \xmark \\
\hline
MoE-Lightning~\cite{cao:25:asplos} & \xmark & \xmark & \xmark & \tickmark & \tickmark & \xmark & \xmark &  \xmark \\
\hline
EdgeMoE~\cite{yi:05:tomc} & \xmark & \tickmark & \xmark & \tickmark & \tickmark & \xmark & \xmark &  \xmark \\
\hline
APEX~\cite{arxiv:25:fan:apex} & \xmark & \xmark & \tickmark & \xmark & \tickmark & \tickmark & \xmark &  \xmark \\
\hline
HeadInfer~\cite{arxiv:25:luo:headinfer} & \xmark  & $-$ & \tickmark & \xmark & \xmark & \tickmark & \xmark &  \xmark \\
\hline
\hline
Pipelined sharding (our work) & \tickmark & \tickmark & \tickmark & \tickmark & \tickmark & \tickmark & \tickmark &  \tickmark \\
\hline
\end{tabular}
\vspace{-0.05in}
\caption{Comparison of Pipelined Sharding with Prior Hybrid Scheduling Approaches.}
\label{tab:hybridschedulers}
\vspace{-0.2in}
\end{table*}

Several researchers have proposed novel solutions to overcome VRAM constraints, including quantization techniques~\cite{neurips:25:hooper:kvquant, aaai:25:zeng:abqllm}, knowledge distillation~\cite{sanh:20:distilbert}, parameter pruning~\cite{muralidharan:24:minitron}, and token compression~\cite{li:23:compress}. Unfortunately, these techniques lower model accuracy and should be the last resort to meet an AI application's performance goals. On the other hand, CPU-GPU offloading or hybrid scheduling techniques are lossless and effectively optimize various aspects of \xlm\ inference. However, no one technique addresses all aspects of VRAM-constrained \xlm\ inference on client systems as shown in Table~\ref{tab:hybridschedulers}.

\begin{figure}[tb]
\centering
\scriptsize
\vspace{-0.05in}
\includegraphics[width=0.994\linewidth]{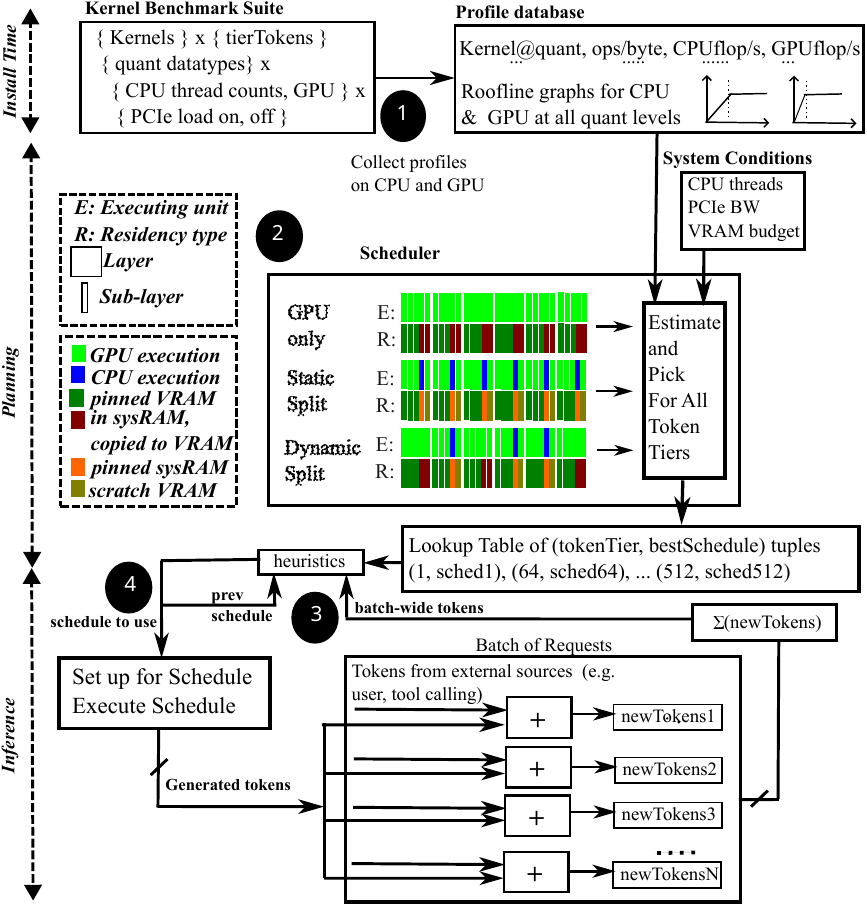}
\vspace{-0.15in}
\caption{Pipelined Sharding Overview. In Step \circled{1}, which runs at install time, we populate a profile database by running a collection of kernels on the CPU and the GPU. In Step \circled{2}, which runs in the planning phase, the scheduler will pin some sub-layers based on priority and generate three possible schedule plans for the remaining sub-layers, namely, a) all of them reside in sysRAM and execute on GPU,  b) some of them are pinned to VRAM scratch and others are sysRAM resident and execute on CPU, and c) make some sysRAM resident sub-layers  to execute on the GPU with just-in-time copies. A timing estimator picks the best schedule for the given system conditions across various token tiers and populates a lookup table with (tokenTier, bestSchedule) tuples. At inference time, in Step \circled{3}, we pick the best schedule for a heuristically determined token tier. Step \circled{4} sets up and executes this schedule for all the requests in the current batch. }
\label{fig:pshardoverall}
\vspace{-0.2in}
\end{figure}

Therefore, to address the above challenge, we present a novel, benchmark-profile-guided inference graph scheduling technique called \emph{pipelined sharding}, illustrated in Figure~\ref{fig:pshardoverall}.
It shards a given \xlm\  at the sub-layer level and schedules these shards on the GPU and the CPU depending on \xlm\ inference conditions (execution phase and context size of in-flight requests), the prevailing system conditions~(CPU thread count, PCIe bandwidth, VRAM budget). 
Our scheduler uses benchmarking-driven timing estimates to select the best among three plans for each one of a set of predetermined token tiers.
While two of the three plans embrace PCIe transfers by overlapping copies of large tensors (e.g., weights, KV cache) with concurrent GPU or CPU compute, the third restricts PCIe copies to just small tensors (e.g., activations). 
During inference, an appropriate tier is chosen and the best schedule for that tier is executed. 
This design helps our algorithm adapt to diverse inference and system conditions and generically handle batches of requests, regardless of the execution phase (context or decode) of individual requests or the batch size.
For VLMs, we combine pipelined sharding with three optimizations, namely, vision tensor offloading to CPU, enabling flash attention for vision encoder~\cite{dao:22:neurips}, and avoiding VRAM allocation overlap between the vision and language parts of the model. These optimizations, which we jointly call \emph{VLMOpt}, though not novel, help bring down vision-related VRAM demand in llama.cpp and make high-resolution VLM inference feasible and efficient across VRAM budgets on client systems. We conclude this section by summarizing our contributions:

1. Pipelined sharding, a novel, benchmark-profile-driven hybrid scheduling technique to optimize LLM inference on client systems, flexibly adapting to various system and inference conditions. With VLMOpt, a llama.cpp implementation of three familiar vision-specific optimizations, pipelined sharding enables efficient, high-resolution VLM inference as well with the same adaptivity. 

2. A C++ implementation of our solution in llama.cpp~\cite{llamacpp} to cater to our C++-first game developer community. We have open-sourced our implementation~(details in Appendix~\ref{sec:artifactappendix}), even as we work to upstream our code to llama.cpp.

3. A rigorous evaluation of pipelined sharding across different dense and MoE LLM models, VRAM budgets, and client systems. On a high-end desktop, for interactive mode inference~(batch size 1), we see an average speedup of 2$\times$ for TTFT  (up to 6.7$\times$)  and 3.7$\times$ for TPS (up to 30$\times$), while for batched inference (batch size $>$ 1), batch-wide TPS improves by \batchavg\ on average (up to \batchmax).

4. An equally rigorous evaluation of the combination of pipelined sharding and VLMOpt for interactive VLMs across client system configurations. Pipelined sharding+VLMOpt reduces VRAM demand for image inference of CR1 by 10$\times$  compared to a vLLM baseline, enabling previously unrunnable high-resolution inference tasks to run successfully in limited VRAM budgets on client systems. 

%% file: background.tex
\section{Background}
\label{sec:background}

In this section, we discuss essential background to set the stage for the rest of the paper.

{\bf Client Systems:} First, it is important for us to clearly define a few minimum requirements for a client system for the purposes of this paper. 
By \emph{client systems}, we specifically mean desktop PCs and laptop computers with a single discrete GPU connected to a CPU via an interconnect such as the Peripheral Component Interconnect Express (PCIe). These client systems have heterogeneous memory, a main or system memory (sysRAM) coupled to the CPU and a video memory (VRAM) attached to the GPU.  
The system's main memory capacity must be greater than a targeted model's total memory requirement, including space to hold model weights and key-value~(KV) cache contents, since our solution relies on being able to pin such data to system memory to achieve good performance. 
Next, we target systems with a minimum main memory bandwidth of 50-100 GBps (consumer laptops and desktops) to  high-end workstations having memory bandwidth around 200-300 GBps. 
While the algorithm itself is GPU-vendor agnostic, our current implementation targets \nvidia\ CUDA-enabled GPUs.

{\bf Products of interest:} Next, we provide a brief overview of the two NVIDIA products we target with pipelined sharding, namely, the IGI SDK~\cite{nvigi} and the CR1 VLM~\cite{nvcosmos, nvidia:25:cr1arxiv}.  The \nvidia\ {\bf IGI~(In-Game Inferencing)} SDK is a  C++ based framework that allows developers to integrate various local AI models into games and other applications, and execute them in-process via CUDA-in-Graphics~(CiG). This allows for optimally scheduled concurrent AI and graphics inside game pipelines. 
The \nvidia\ {\bf CR1~(Cosmos-Reason1)} is a reasoning VLM designed for physical AI applications. It processes images at native resolution, uses chain-of-thought reasoning to interpret physical scenes, and generates natural language for embodied decisions (e.g., selecting the next action). Its native-resolution processing makes it VRAM-intensive on client hardware, motivating the optimizations in Section~\ref{sec:vlmopt}.

{\bf Autoregressive models:} Modern decoder-only {\xlm}s tokenize inputs into embeddings, which are then fed through a stack of serially-executing transformer layers. Each layer contains an attention sub-layer (multi-head or group query) followed by a feed-forward network~(FFN) block or sub-layer. Dense models use a single FFN, while MoE models use a gating function to route each token to a subset of expert FFNs and combine their weighted outputs.
In both cases, after all layers are processed, the output is normalized and sent to a linear layer to output one token.

An LLM  starts with an input token sequence of a certain length~(ISL). It first performs a context (or prefill or prompt) phase where it processes all input tokens concurrently in a compute-intensive pass that populates the attention scores across the full input sequence.
This  key~(K) and value~(V) inputs to attention blocks get cached in a structure called the KV cache. 
It then enters a memory-bandwidth-limited decode phase, producing one token per iteration using cached KV values.
A model's context is defined as the concatenation of the input sequence and the space reserved for the maximum expected output token count. 

{\bf Interactive TPS rates:} For {\xlm}s to be practically useful, we argue that their TPS rates must be good enough to keep up with human reading rates. Average human reading rates (English) are 183 words per minute~(WPM) for oral reading and 238 WPM for silent reading~\cite{brysbaert:19:jml}. Using the rule of thumb of 0.75 words per token~\cite{openaitokens}, this translates to approximately 4-5 TPS. User preferences align with this threshold too~\cite{tps}. As we will see in Section~\ref{sec:eval}, pipelined sharding not only enables large models to fit in smaller VRAM budgets, but also elevates their performance to above the 5 TPS threshold, enabling deployment in demanding client AI applications.

%% file: related.tex
\section{Prior Work}
\label{sec:related}

In this section, we compare our work qualitatively with prior work on VRAM-constrained single-GPU inference.

Sharding splits a large \xlm's computation graph into sub-graphs or shards such that no shard's VRAM requirements exceed the specified VRAM budget. It then dynamically maps shards serially onto the GPU in topological sort order, copying necessary shard input data (weights as well as activations from prior shards) and performing inference computation to produce one token of output. This process is repeated in a loop to produce the full output. This is a basic technique with the singular focus of getting large models to successfully run on VRAM-constrained systems~\cite{lepikhin:20:gshardarxiv}. 
Llama.cpp's static layer partitioning shards an \xlm\ at layer boundaries and maps them to the CPU or the GPU based on a manually specified command-line parameter~\cite{llamacpp}.

DeepSpeed's ZeRO-Inference optimization uses prefetching of layer weights and large batch sizes to amortize the cost of PCIe copies and achieve good inference throughput~\cite{sc:22:aminabadi:deepspeed}. 
Elastic Pipelining breaks a model into multiple sub-layer shards, compiles them with different bit-widths, profiles each shard's relevance to overall accuracy, and then during inference, picks the most appropriate shard to execute, balancing memory use, latency, and accuracy on mobile devices~\cite{guo:23:asplos}. This is a lossy technique to minimize memory demand in mobile devices. 
FlexGen uses two strategies - it uses 4-bit quantization to reduce model size and then uses integer linear programming to find an optimal schedule to maximize throughput for large batch size scenarios~\cite{icml:23:sheng:flexgen}. 

PowerInfer categorizes neurons as hot or cold depending on how frequently they get activated, and uses an effective predictor to speculatively execute hot neurons on the GPU and cold neurons on the CPU~\cite{sosp:24:song:powerinfer}. 
Its success is dependent on high-sparsity activation functions used in older models. PowerInfer-2 uses a profile-driven approach to determine hot and cold neurons,  orchestrates neuron transfer from flash storage to RAM on mobile devices, and maps neurons to the CPU or neural processing unit~(NPU) of mobile devices based on their computational density~\cite{xue:24:powerinfer2}. In contrast, our approach focuses on discrete GPU setups where it has to contend with PCIe transfer costs. Our algorithm works well across LLM phases and batch sizes.

Finally, we compare our technique succinctly with prior hybrid scheduling techniques in Table~\ref{tab:hybridschedulers}.  As seen from that table and to the best of our knowledge, pipelined sharding is a unique benchmark-profile-based hybrid scheduling solution that can adapt to the inference phase, batch size, context size, and system conditions, and works well for both dense and MoE LLMs.

%% file: pipeshard.tex
\section{Pipelined Sharding}
\label{sec:pipeshard}
\makeatletter
\newcommand{\removelatexerror}{\let\@latex@error\@gobble}
\makeatother

\newcommand{\includeForAlgo}[0]{
\scriptsize
\SetAlgoLined\DontPrintSemicolon
   \SetKwFunction{CalculateVRAMRequired}{CalculateVRAMRequired}
   \SetKwFunction{PinShardsToGPU}{PinShardsToGPU}
   \SetKwFunction{ReserveVRAM}{ReserveVRAM}
   \SetKwFunction{SortByPriority}{SortByPriority}
   \SetKwFunction{PeekHead}{PeekHead}
   \SetKwFunction{PopHead}{PopHead}
   \SetKwFunction{ShardIntoSubLayers}{ShardIntoSubLayers}
   \SetKwFunction{ShardIntoLayers}{ShardIntoLayers}
   \SetKwFunction{CopyAndGPUExecute}{CopyAndGPUExecute}
   \SetKwFunction{DecideScratchVRAMBudget}{DecideScratchVRAMBudget}
   \SetKwFunction{Mark}{Mark}
   \SetKwFunction{IsContextPhase}{IsContextPhase}
   \SetKwFunction{TransformGraph}{TransformGraph}
   \SetKwFunction{RunInTopologicalOrder}{RunInTopologicalOrder}
   \SetKwFunction{MarkForContextPhase}{MarkForContextPhase}
   \SetKwFunction{PlanningPhase}{PlanningPhase}
  \SetKwFunction{HeuristicSplit}{HeuristicSplit}
  \SetKwFunction{RunBenchmarkingOnce}{RunBenchmarkingOnce}
  \SetKwFunction{PickTier}{PickTier}
  \SetKwFunction{SetupForSched}{SetupForSched}
   \SetKwProg{myproc}{Procedure}{}{}
   \SetKwInOut{Input}{I}
   \SetKwInOut{Output}{O}
   \SetCommentSty{mycommfont}
}

Pipelined sharding takes inspiration from multiple past ideas to arrive at an efficient technique for both interactive~(batch size 1) and batched processing (batch size $>$ 1) of LLM inference on client systems.
It is a fully automatic hybrid scheduling technique that breaks an input inference graph at the sub-layer level into sub-graphs called shards and picks the most appropriate schedule for these shards for the given system and inference conditions, from a small number of pre-generated schedules.
For each shard, it determines its memory residency (i.e. sysRAM or VRAM) and execution backend (CPU or GPU), ensuring, as far as possible that PCIe copies of tensors get pipelined with and hidden under concurrent GPU or CPU compute work.  
Our scheduler pre-assigns backends for the majority of sub-layer components to facilitate semantically aware graph cuts. These cuts occur at meaningful functional boundaries (e.g., between attention and FFN sub-layers), enabling precise control over which operations execute together.
The default llama.cpp scheduler, which runs after ours, honors these assignments. For tensors and ops that our scheduler did not pre-assign, the default scheduler places them such that cross-backend communication is minimized. 
Algorithm~\ref{alg:pshard} shows the high-level procedure of pipelined sharding. It operates in three phases, namely, install, planning, and inference, as shown in Figure~\ref{fig:pshardoverall}. 
We discuss the three phases in turn below.

{\bf Install Phase:} To guide scheduling decisions in our algorithm's profile-guided planning phase, we first profile a suite of benchmark kernels during llama.cpp installation.  An offline benchmarking run collects CPU and GPU floating point operations per second~(FLOPS) metrics for  kernels (tensor operations) relevant for modern autoregressive transformers.  It is a generic benchmark comprising kernels for matrix multiply (matmul), group query attention (GQA), multi-head attention (MHA), mixture-of-experts routing (MoE), and element-wise operations. It runs these kernels across a range of tensor sizes, quantized datatypes, context sizes, and number of KV heads.
For CPU profiling, we vary thread counts and measure performance, both standalone and under concurrent PCIe traffic to capture memory controller contention. For each configuration, we average over multiple iterations to reduce measurement noise. For GPU profiling, we launch ten asynchronous kernel calls and measure their total completion time.
This GPU-specific strategy helps measure wider GPUs' ability to absorb more concurrent work for narrow kernels.  On narrow GPUs or for wide kernels, these calls may serialize naturally.  Since we measure the total time for all ten independent calls as well as their cumulative operation count, the resulting FLOPS value implicitly captures the concurrency offered by the underlying GPU for a given kernel.
With the above setup, a one-time run of the benchmark (takes about 15 minutes) gives us a rich kernel-level performance profile for both the CPU (across thread counts, with and without PCIe interference) and the GPU. The generated profile takes up a modest 170KB of disk space. We will discuss how this profile is used for schedule cost estimation later in this section.

\underline{Profile Robustness:} Our profiler design philosophy is to populate the profile database with profiles obtained under an almost universal set of possible system conditions (e.g. varying CPU thread count, PCIe BW, PCIe-main memory contention, etc.) just once at install time and then look up this database based on actual prevailing system conditions at planning time (i.e. during xLM invocation). Specifically, we use the CPU thread count passed in via the llama.cpp command line and PCIe BW demand of a schedule to model main memory bandwidth contention to retrieve appropriate entries from the database to estimate schedule costs. 
Profiles remain valid under most system condition changes. We only anticipate having to re-generate profiles when core GPU software libraries are updated, which may impact GPU timing estimates for our kernels. Though not yet implemented, we expect such library changes can be automatically detected to trigger re-profiling and update the database.
Thermals are not handled currently, but the same design philosophy can be extended to adapt to dynamic power state changes (i.e., \emph{a priori} profiling under all power states, followed by planning time lookup with prevalent power state).

\newcommand{\chosenTier}[0]{$t$}

{\bf Planning Phase:}  Pipelined sharding's \emph{token tier design} allows it to elegantly create, maintain, and run inference schedules optimized for a set of supported ``new token'' tiers. 
Here ``new token'' refers to the total number of new tokens that must be processed in one inference iteration for a batch of requests. This token count determines the sizes of various tensors involved in inference. 
Higher batch-wide new token counts (one or more context phase requests or a large number of decode phase requests) will cause a batch's inference performance to be compute bound. 
Lower new token counts (no context phase requests and few decode phase requests) mean reduced compute demand (thanks to the use of pre-populated KV caches), making performance memory bound.  
As token counts vary from iteration to iteration, tensor sizes and scheduling tradeoffs will vary too. The rationale to support multiple token tiers is to optimize schedules based on tensor sizes. 
Thus, in the planning phase, for each of a set of supported new token tiers, we pre-generate three different schedule plans, pick the best among them, and save it in a token tier table. 

For a given tier, we first shard the input graph at the sub-layer level for finer-grained control over tensor placement. 
Inspired by EdgeMoE~\cite{yi:05:tomc}, we found it effective to prioritize attention sub-layers over FFNs for VRAM allocation in both MoE and dense models. We extend this idea to include more sub-layer types for prioritized VRAM allocation and found the following decreasing priority order useful in practice: attention, KV cache, FFN, and outputs.  We partition the specified VRAM budget into pinnable and scratch areas. We pin as many high-priority sub-layers as possible in the pinnable area to avoid repeated PCIe transfers during inference. Our token tier approach leverages the smaller intermediate tensor sizes at lower token tiers to pin more sub-layers to VRAM than llama.cpp's static layer partitioning approach.
For the remaining sub-layers residing in system memory, we evaluate three distinct plans.

\begin{algorithm}[t]
\vspace{-0.1in}
\includeForAlgo\
\BlankLine
\BlankLine
\Input{B : vramBudget, G : graph}
\nl\myproc{PipelinedSharding{}} {
\tcc{Install phase.}
\nl Profile $\leftarrow$ \RunBenchmarkingOnce{}\;
\tcc{Planning phase.}
\nl $P$ $\leftarrow$ [attn, kvcache, ffn, outs]\;
\nl \For {Tier $\in$ {1, 4, 16, 32, ..., 16K}} {
\nl \PlanningPhase{B, G, P, Profile, Tier}}
\tcc{Inference phase}
\nl $\mathcal{T}$ $\leftarrow$ \PickTier{batchWideTokenCount}\;
\nl \SetupForSched{sched[$\mathcal{T}$]}\;
\nl \RunInTopologicalOrder{sched[$\mathcal{T}$]}}
\BlankLine
\BlankLine
\Input{B : vramBudget, G : graph, P : shardPriority, profile : profileInfo, tier : int}
\nl\myproc{PlanningPhase{}} {
\nl  SL$_\text{all}$ $\leftarrow$ \ShardIntoSubLayers{G}\;
\nl  B$_\text{scratch}$ $\leftarrow$ \DecideScratchVRAMBudget{B, G, SL$_\text{all}$}\;
\nl B$_\text{pinned}$ $\leftarrow$ B $-$ B$_\text{scratch}$\;
\nl  SL$_\text{rem}$ $\leftarrow$ \PinShardsToGPU{B$_\text{pinned}$, SL$_\text{all}$, P}\;
\nl (SL$_\text{cpu}$, SL$_\text{gpu}$) $\leftarrow$ \HeuristicSplit{P, profile, SL$_\text{rem}$}\;
\nl   \For {sl $\in$ SL$_\text{cpu}$} {   
\nl  \Mark{sl, Residency=SysMem, Exec=CPU}}
\nl   \For {sl $\in$ SL$_\text{gpu}$} {   
\nl  \Mark{sl, Residency=SysMem, Exec=GPU}}
\tcc{add to schedule table for this $tier$}
\nl  sched[tier] += SL$_\text{all}$\;
}
\BlankLine
\BlankLine
\Input{B : vramBudget, S : shardSet, P : shardPriority}
\Output{B$_\text{rem}$ : remainingVRAM, S$_\text{rem}$ : remainingShards}
\nl\myproc{PinShardsToGPU{}} { 
\nl S$_\text{sorted}$ $\leftarrow$ \SortByPriority(S, P)\;
\tcc{Fit as many shards as possible in $B$.}
\tcc{Return remaining VRAM and shards} 
\nl \KwRet{B$_\text{remaining}$, S$_\text{remaining}$}}\;
\BlankLine
\caption{Pipelined Sharding.}
\label{alg:pshard}
\vspace{-0.2in}
\end{algorithm}

\underline{Plan One (\plangpuonly)}: 
All remaining (i.e. unpinned) sub-layers execute on the GPU. Their weights must be copied before each use to a VRAM-scratch-based double-buffer. 
A batch with high new token count will prefer this schedule as its high compute demand can hide PCIe transfer latency. On the other hand, for a batch with a low new token count, PCIe bandwidth becomes the bottleneck, as transfers cannot overlap with GPU work. Here, the CPU is not used as a compute resource.

\underline{Plan Two (\planstatic)}: We permanently partition the unpinned sub-layers between the GPU and the CPU based on priority. High-priority sub-layers get pinned to VRAM scratch and run on the GPU, while low-priority ones stay in system memory and run on the CPU. This plan  transfers only intermediate outputs between the CPU and the GPU, avoiding repeated weight transfers. If sufficient CPU threads are available, main memory bandwidth becomes the limiting factor for CPU-assigned sub-layers.

\underline{Plan Three (\plandynamic)}: A hybrid approach where fewer sub-layers run on the CPU compared to Plan Two, while more run on the GPU by time-sharing a VRAM-scratch-based double-buffer. Streaming of weights from sysRAM for a GPU-mapped sub-layer overlaps with a concurrently executing prior CPU sub-layer. A key point to note is that CPU computation and PCIe transfers share the same memory controller, creating sysRAM bandwidth contention. When PCIe is active, CPU gets less bandwidth than it would in isolation (and vice versa). This means both operations slow down compared to running alone, but total throughput can still be higher. Fewer CPU threads will suffice to saturate the reduced per-operation bandwidth.

Each of these plans may be attractive under different situations. For example, if very few CPU threads are available due to other concurrently running applications (e.g. video games), the CPU will not be able to saturate sysRAM bandwidth. 
Here, GPU offloading makes sense despite PCIe overhead, favoring Plan \plangpuonly\ or Plan \plandynamic\ based on CPU thread availability. 
On the other hand, with ample CPU threads and large tensors to transfer (e.g. KV caches for long contexts), Plan \planstatic\ avoids repeated copy of such tensors across PCIe for every decode iteration. The three plans represent the fundamental ways to balance the CPU-GPU-PCIe resource
triangle under VRAM constraints. \plangpuonly{} maximizes GPU utilization when PCIe bandwidth is adequate. \planstatic{} minimizes PCIe traffic
when CPU threads are plentiful. \plandynamic{} overlaps CPU work with GPU weight streaming when both resources can be utilized
concurrently. Our infrastructure is designed for extensibility: schedules are parameterized through a configuration table that defines tensor placement and sharding behavior, allowing new schedules to be added without modifying the core planning logic. 
{\tt HeuristicSplit()} implements the above functionality, generating all three plans and selecting the best one using profile-based timing estimation, which we describe next.

{\it Profiler-based timing estimation for schedule plans:}  For a given plan, for each constituent kernel of each sub-layer mapped to the CPU, we compute its CPU-limited time as follows. 
For a kernel $K$, we first check if it is present in our profile database (exact match on op name, quantization level, available CPU thread count, and op dimensions). If present, we retrieve its FLOPS value from the database and divide the number of floating point operations by the retrieved FLOPS value to get the CPU-limited time. 
Otherwise, if there is a partial match on the op name, quantization level, and CPU thread count, but only the op dimensions differ, we attempt a nearest neighbor search (in the space of op dimensions) over the subset of such partial-match kernels and pick the one with the shortest distance to $K$. 
We then construct a roofline graph with this chosen benchmark kernel~\cite{williams:09:cacm, yuan:24:llmroofline}. Depending on where $K$'s arithmetic intensity lands in this roofline graph, we identify $K$ as compute bound or memory bound. If it is compute bound, we divide $K$'s floating point operation count by the FLOPS roofline value to get its timing estimate. On the other hand, if $K$ is memory bound we divide $K$'s byte count by the memory bandwidth to get its timing estimate.  
If $K$ fails to get a partial match (usually metadata operations like RESHAPE, VIEW with negligible computational cost), we skip it in our timing estimate.
Should new, computationally significant kernels show up in newer {\xlm}s, the benchmark suite will need to be updated to represent them suitably. 
For 11,126 ops across our test LLMs, 24\% get an exact match, 32\% a partial match, and the remaining 44\% are skipped.

Next, we add up the time estimates for all such kernels across all sub-layers to get the total CPU-limited time for a particular plan. 
We use a similar procedure to estimate GPU-limited time for a plan by applying the above procedure to sub-layers mapped to the GPU. 
Each shard's estimated time accounts for concurrent PCIe transfers, if any. While a shard executes, we may transfer: (1) weights and KV cache for the next shard to VRAM (host-to-device), and/or (2) KV cache from the previous shard back to sysRAM (device-to-host). The shard's estimated time is the maximum of its compute time (CPU or GPU, as appropriate) and these concurrent PCIe transfer times. If no concurrent transfers occur, the shard's time is simply its compute time. We sum all constituent shard times to get a plan's total estimated time and then select the plan with the shortest time as a given tier's plan. 

{\bf Inference Phase:}  Recall that the planning phase populates a ``new token'' tier table with schedules optimized for individual token tiers. 
We look up this table with the batch-wide new token count for every inference iteration. 
To determine which tiers to support, we note that a batch may have one or more requests and individual requests may be in their respective context or decode phases.
A request in the context phase typically contributes a large number of new tokens to the batch-wide new token count, while a request in the decode phase contributes just one new token. 
Since new token counts can vary with input prompts, inference phases, or batch sizes, it becomes necessary to support a broad range of token tiers to handle various scenarios. 
Thus, the set of supported token tiers, $\mathbb{J}= \{ 1, 4, 16, 32, 64, 512, 1K, 2K, 4K, 8K, 16K \}$.
Then, for every inference iteration, we automatically pick a tier $\mathcal{T}$ that minimizes the product $\lceil\frac{\text{batchwideNewTokens}}{\text{tierSizeInTokens[\chosenTier]}}\rceil * \text{\scriptsize estimatedSchedTime[\chosenTier]}, \forall t \in \mathbb{J}$, where the first factor is the number of iterations through \chosenTier's schedule and the second is the latency of one trip through that schedule, and set up for and execute $\mathcal{T}$'s schedule. 
For context phase requests, $\mathcal{T}$ serves as the optimal chunk size for chunked prefills~\cite{agrawal:24:osdi}.

%% file: vlmopt.tex
\section{VLMOpt}
\label{sec:vlmopt}

VLMs require careful VRAM management for vision and language processing. While pipelined sharding handles the language part, 
this section details our engineering contributions to llama.cpp's vision encoder to allow for efficient high-resolution image VLM inference. We implement complementary VRAM-reduction optimizations, jointly called \emph{VLMOpt}, directly inside llama.cpp's multimodal path so that downstream adapters benefit from these optimizations without any extra changes.  In this paper, though we focus specifically on the Qwen2.5-VL family~\cite{bai:25:tr} and their derivatives such as CR1, these  optimizations are generic and can be enabled for other VLMs as well.

The first optimization, Vision Tensor Offload, modifies llama.cpp's CLIP/vision model loading path so that all vision model's weight tensors are allocated and pinned on the CPU. We create per-buffer allocation contexts on the host and bind tensor data to them. Weight tensors stream from CPU to GPU at execution time. The rest of the scheduling and execution remain unchanged.

The second optimization addresses a fundamental problem in llama.cpp's Qwen2.5-VL vision encoder which prevents it from handling high-resolution images.  Specifically, the self-attention logic in the vision encoder creates O(N$^2$) score tensors KQ and associated intermediates, where N is the number of vision tokens.  For high-resolution images, KQ alone can exceed several gigabytes, making inference impractical on single client GPUs.  We address this by using llama.cpp's implementation of FlashAttention~\cite{dao:22:neurips}, hitherto used only for language decoding, in its Qwen2.5 VL vision encoder as well.
On top of this, we tile query (Q) tensors to control sizes of FlashAttention's input tensors.
Though this Q-chunking hurts performance by breaking FlashAttention's cache-friendly working sets and requiring concatenation of partial outputs, it enables us to arbitrarily reduce the vision VRAM usage for high-resolution images like 1440p to under 2 GB.

The third optimization restructures the llama.cpp code to ensure that vision encoding is completed and all its GPU resource allocations are freed prior to the language context initialization, which happens just in time after all the multimodal tokens are arranged and prior to the first language decode call.  This serialization ensures peak VRAM is the greater of the VRAM demand of the vision and language parts, as opposed to the baseline where it equaled their sum.

%% file: methodology.tex
\section{Methodology}
\label{sec:methodology}

\begin{table}[htb]
\vspace{-0.1in}
\centering
\scriptsize
\begin{tabular}{llr}
\hline
{\bf Short Name} & {\bf Model Name} & {\bf Size on disk} \\ 
\hline
\multicolumn{3}{c}{{\em IGI SDK: Two dense LLMs and one VLM (CLIP enc/dense dec)}} \\
\hline
nemo4b & mistral-nemo-minitron-4b-128k-instruct-f16 & 7.7 GB \\
nemo8b & mistral-nemo-minitron-8b-128k-instruct-f16 & 15.7 GB \\
vnemo4b & nemotron-vision-4b-instruct-f16  & 8.4 GB\\
\hline
\multicolumn{3}{c}{{\em Two forward-looking MoE LLMs}} \\
\hline
qwen30b & Qwen3-30B-A3B-Instruct-2507-q4 &  16.4 GB \\
qwen235b & Qwen3-235B-A22B-Instruct-2507-q2\_k & 77.0 GB\\
\hline
\multicolumn{3}{c}{{\em Cosmos-Reason1 reasoning VLM (CLIP enc/dense dec)}} \\
\hline
cr1 & Cosmos-Reason1 & 15.4 GB\\
\hline
\end{tabular}
\vspace{-0.1in}
\caption{Model details.}
\label{tab:models}
\vspace{-0.1in}
\end{table}

We evaluate our solution on six different models - two LLMs and one VLM  
from the IGI SDK, the CR1 VLM, and two powerful MoE LLMs. The details of these models are given in Table~\ref{tab:models}. We chose these models to demonstrate that our optimization is effective across both dense and MoE LLMs.
We have implemented our solution in the llama.cpp framework (branch 6097).  
Since our optimizations preserve model accuracy and do not change the operations executed, we focus on measuring runtime performance using controlled test inputs rather than standard benchmark suites. For our LLMs, we use our own test inputs to precisely control context size and output sequence length, ensuring repeatable TTFT and TPS measurements.
For VLMs, we quantify TTFT and TPS (for language), and vision encoding time. 
Since \app{vnemo4b} operates only on 336x336 images internally, its vision token count is resolution independent and so we test it with just a single input image.
On the other hand, since \model{cr1} processes images at their native resolution, its vision token count increases with input image resolution. We test it with images  at multiple resolutions~(480p, 720p, 1080p, 1440p) to show that it benefits from our  VRAM-aware flash attention optimization for vision encoding.

\begin{minipage}[htb]{\linewidth}
\vspace{-0.1in}
\scriptsize
\begin{table}[H]
\scriptsize
\centering
\begin{tabular}{p{0.15in}p{0.3in}p{0.2in}p{0.5in}p{0.2in}p{0.4in}p{0.4in}}
\hline
{\bf Name} & {\bf GPU} & {\bf VRAM (GB)} & {\bf cores (CPU)} &  {\bf RAM (GB)} & {\bf Mem BW (GBps)} & {\bf PCIe BW (GBps)} \\
\hline
\machine{cli1} & 3500  &  12 & 16 (Ultra7) & 64 & 119.5 & 13 (16\footnote{\label{firstlabel}theoretical peak value}) \\
\machine{cli2} & 5070 TI & 16 &  8 (Ryzen7) & 128 & 57.6  & 50 (64\footref{firstlabel}) \\
\machine{cli3} & 5090 & 32 &  16 (EPYC) & 256 & 153.6 & 50 (64\footref{firstlabel}) \\
\hline
\end{tabular}
\vspace{-0.2in}
\caption{Client systems used for evaluation.}
\label{tab:machines}
\vspace{-0.1in}
\end{table}
\vspace{-0.1in}
\end{minipage}

Details of our test client systems, which include a laptop~(\machine{cli1}) a typical client desktop~(\machine{cli2}), and a high-end desktop~(\machine{cli3}), are provided in Table~\ref{tab:machines}.
All our client systems run the Windows 11 operating system.

To the best of our knowledge, none of the currently popular inference frameworks can take VRAM budget as a command-line parameter and optimize for that budget. Llama.cpp requires users to manually specify the number of GPU layers (\emph{-ngl} parameter) to offload. Since actual VRAM usage depends on runtime factors like intermediate tensor sizes, batch size, and allocator behavior, it cannot be determined statically. Finding the optimal \emph{ngl} value for a given VRAM budget therefore requires iterative trial-and-error: load the model with a candidate \emph{ngl} value, run inference to trigger graph generation and allocation, measure actual VRAM usage, adjust \emph{ngl}, and repeat until the budget is met without exceeding it. For our baselines, we performed this search process for each model-budget combination to find the maximal number of layers that fit within the given VRAM budget. We use these aggressive baselines, called \emph{llama-cpp-baseline}, for fair performance comparisons. In contrast, our solution automatically determines optimal placement given only a VRAM budget, eliminating this manual tuning overhead for users.
The baseline for CR1 is the vLLM framework, which we run on Windows Subsystem for Linux~(WSL), for the max VRAM budget and for lower budgets, we use \emph{llama-cpp-baseline}.

Finally, a short note on metrics. In addition to TTFT and TPS, we also calculate end-to-end-latency~(E2EL) as follows. We measure TPS over 100+ output tokens and then calculate E2EL for a standard  output sequence length of 100 tokens as \emph{TTFT + 100 / TPS} for LLMs and \emph{VisionEncTime + TTFT + 100 / TPS}  for VLMs.

%% file: eval.tex
\section{Evaluation}
\label{sec:eval}
\begin{table*}[htb]
\vspace{-0.05in}
\centering
\scriptsize
\begin{tabular}{c|c|rrrrrrrr|rrrrrrrr}
  & Ctx & \multicolumn{8}{c}{TPS  across VRAM budgets} & \multicolumn{8}{c}{TTFT in \emph{seconds} across VRAM budgets}  \\
 Model & Size & 2G & 4G & 6G & 8G	& 12G &16G	& 24G & 32G & 2G & 4G & 6G & 8G & 12G &16G	& 24G & 32G \\
 \hline
\hline
nemo- & 1K &	14.8	& 20.1 & 33.1 &	66.1 & 116.9 & 116.9 & 116.9	 & 116.9 & 0.2	& 0.2	&	0.2	& 0.1 & 0.1 & 0.1 & 0.1 & 0.1\\
 4b	   & 4K &   12.3   & 18.4 & 28.8  &  63.1 & 108.7 & 108.7 & 108.4   & 108.4 & 0.5	& 0.4	&	0.4	& 0.4 & 0.4 & 0.4 & 0.4 & 0.4 \\
	    & 16K &  5.3   & 12.2	 & 18.7  & 29.5   & 87.9	 &  87.9  & 87.9   & 87.9  & 2.7	& 2.5	&	2.4	& 2.2 & 2.0 & 2.0 & 2.0 & 2.0 \\
	    & 64K & 	1.3   &  1.6	& 2.0     & 3.2	    & 15.0   & 35.8	  & 50.3  & 50.3 & 28.8 & 27.7  &   26.0	& 24.7 & 22.1 & 19.1 &	17.2	& 17.2 \\
\hline
nemo- & 1K &	7.6 & 8.6 & 10.0 & 12.4 & 21.6 & 70.0 & 70.0 & 70.1  & 0.4	& 0.3 & 0.3 & 0.3 & 0.2 & 0.2	& 0.2 & 0.2\\
8b        & 4K &   6.4 & 8.2 &  9.5   & 11.6  & 20.3 & 64.4 & 66.7 & 66.6 & 0.8 &  0.8 & 	0.7	& 0.7 & 0.7 & 0.6    & 0.6	  &  0.6 \\
	    & 16K &  3.3 & 4.8 &	8.1 & 	9.5 & 14.5  & 28.5 & 56.9 & 56.7  & 4.1 & 3.9  & 3.8    & 3.7	 & 3.4  & 3.1   & 3.0  & 2.9 \\
	    & 64K & 	1.0 & 	1.1 & 1.4 & 1.7    & 3.8	& 8.8	 & 25.2 & 36.5  &  53.8	 & 35.6 &	34.6	& 33.3 & 31.2 & 28.9 & 24.6 & 22.6\\
\hline
qwen- & 1K & 25.7	& 26.2 & 29.6 & 32.1 & 36.5 & 47.8	 & 158.1 & 158.6 & 0.6 &  0.6 & 0.5 & 0.5 & 0.4 & 0.3 & 0.3 & 0.3 \\
30b      & 4K &  25.6	& 26.7 & 28.3 & 31.7 & 37.0 & 43.4	 & 141.3 & 141.3 & 1.5	 & 1.5 & 1.4 & 1.3 & 1.2 & 1.0 &  1.0 & 1.0 \\
                & 16K & 20.4  & 25.1 & 27.4 & 28.7 & 33.3 & 39.1 & 101.5 & 100.9  & 6.8 & 6.5 & 6.2 & 5.8 & 5.3 & 4.6  & 4.1 & 4.1 \\
                & 64K & 4.7    & 6.7	   &  11.8 & 21.1 & 24.2 & 27.2 & 60.6   & 60.6    & 42.1 & 41.3  & 39.7 & 38.5 & 35.9 & 33.1 & 26.7 & 26.6\\
\hline
qwen- & 1K & 7.7 & 8.7 & 9.0 & 9.1 & 9.4 & 9.7 & 10.6 & 11.5 & 2.5 & 2.4 & 2.4 & 2.4	 & 2.3 & 2.3 & 2.1 & 2.0 \\
235b     & 4K & 7.5 & 8.3 & 9.0 & 9.0 & 9.3 & 9.7 & 10.4	& 11.1 & 6.7 & 6.7 & 6.7 & 6.6 & 6.5 & 6.4 & 6.1	& 5.9 \\
                  & 16K & 5.2	& 6.7	& 7.6 & 8.6	& 8.9	& 9.3	& 9.9	& 10.9 & 28.8 &	28.8	& 28.6 & 28.4 & 27.9 & 27.5 & 26.5 & 25.7\\
                  & 64K & 2.0	& 2.1	& 2.4	& 2.8	& 4.1	& 7.5	& 8.2	& 8.7   & 148.1 & 148.0 & 147.2	& 146.7 & 145.2 & 143.8 & 140.7 & 138.2 \\
\hline
\end{tabular}
\vspace{-0.1in}
\caption{TPS and TTFT from pipelined sharding  on \machine{cli3}. Column headers are VRAM budgets.}
\label{tab:tps}
\vspace{-0.15in}
\end{table*}

In this section, we perform a detailed evaluation of pipelined sharding and VLMOpt under different test conditions. 
We demonstrate that our solution is effective in fitting large {\xlm}s in any VRAM budget on client systems and does so efficiently. 
We also perform a few relevant sensitivity studies to demonstrate the adaptivity of our solution. Note, our VRAM budget specification is always in multiples of 1,000 MB, not 1 GB (i.e., 1,024 MB). To highlight this, we suffix VRAM budgets with just `G' and not `GB'. 

\begin{figure}[htb]
\centering
\scriptsize
\includegraphics[width=\linewidth]{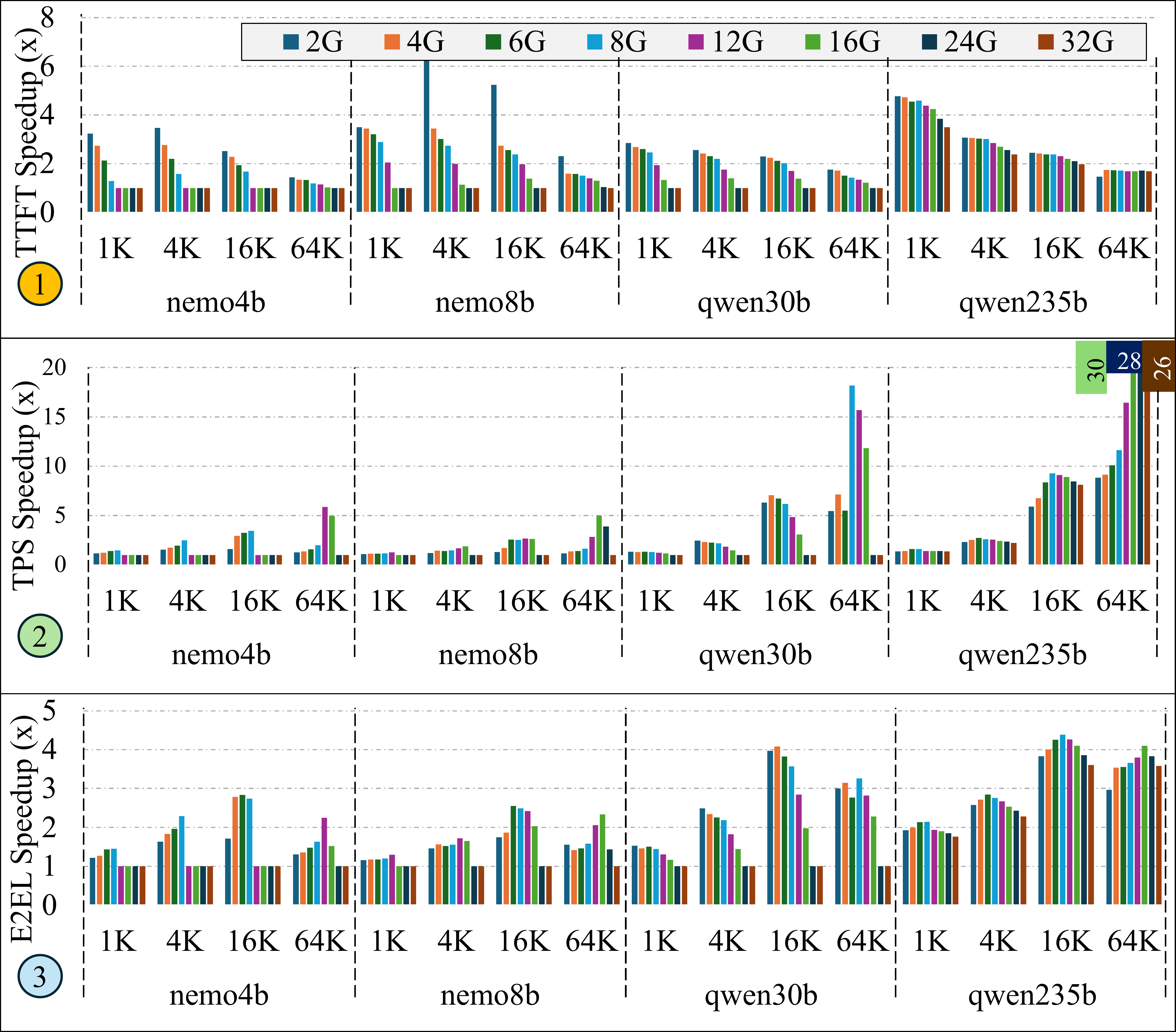}
\vspace{-0.2in}
\caption{Speedups from pipelined sharding for LLMs on \machine{cli3}, relative to \emph{llama-cpp-baseline}. Each chart shows speedups across eight VRAM budgets (2G to 32G) for four models at four context sizes (1K, 4K, 16K, 64K). (1) Time-to-first-token (TTFT) speedups average 2$\times$ (max 6.7$\times$). (2) Tokens-per-second (TPS) speedups average 3.7$\times$, reaching up to 30$\times$ for qwen235b at 64K context (truncated values labeled above). (3) End-to-end latency (E2EL) speedups average 2$\times$ (max 4.3$\times$).}
\label{fig:bestresults}
\vspace{-0.1in}
\end{figure}

{\bf LLM Performance:}  Table~\ref{tab:tps} quantifies raw pipelined sharding performance (TPS and TTFT) for our test models on \machine{cli3}. Unless otherwise stated, across all our experiments, batch size is 1 and llama.cpp is made to use as many CPU threads as physical cores (i.e., 16 on \machine{cli1} and cli3, 8 on \machine{cli2}).  A key observation from Table~\ref{tab:tps} is that our technique enables large models to achieve interactive performance ($\geq$ 5 TPS) even at minimal VRAM budgets. At just 2G VRAM, \model{qwen235b} (77GB on disk, a 39$\times$ reduction) delivers 7.7 TPS for 1K context and remains interactive (5.2 TPS) even at 16K context. Similarly, \model{qwen30b} achieves 20-26 TPS at 2G VRAM for contexts up to 16K, demonstrating that massive models can run efficiently on severely VRAM constrained systems. At 4G VRAM, most models exceed the 5 TPS threshold for contexts up to 16K, with \model{qwen235b} delivering 6.7 TPS at 16K context despite using only 5\% of its 77GB disk footprint in VRAM. 

Figure~\ref{fig:bestresults} shows speedups for \machine{cli3} relative to \emph{llama-cpp-baseline} (with key results summarized in its caption). Notice that speedups vary with context size and VRAM budget. As context size increases, KV cache bloats and overall time spent on the context phase grows quadratically, dominating E2EL. This quadratic growth, combined with better baseline GPU utilization at higher VRAM budgets, reduces speedups at larger contexts. For instance, \model{qwen235b} records its lowest TTFT speedup (1.68$\times$) at 64K context with 32G VRAM.  Exceptional TPS speedups (up to 30$\times$) for large contexts are possible because our solution considers the KV cache too as part of its shard scheduling decisions, and thus gracefully adapts to different context sizes. 

\begin{table}[htb]
\centering
\scriptsize
\begin{tabular}{c|c|rr|rr}
Model  & Ctx Size & \multicolumn{2}{c}{cli2 best metrics} & \multicolumn{2}{c}{cli1 best metrics}  \\
 &  & TPS & TTFT (s) & TPS & TTFT (s)\\
 \hline
\hline
nemo4b & 1K & 83.1 &  0.12 & 34.9 & 0.29 \\
              & 4K & 76.8 & 0.45 & 32.3 & 1.19 \\
              & 16K & 58.8 & 2.44 &  24.7 & 6.63\\
              & 64K & 14.4 & 30.22 & 6.0 & 85.49\\
\hline
nemo8b & 1K & 22.9 & 0.23  &	10.7 & 0.76\\
           &  4K & 22.3 & 0.78 & 9.7 & 2.37 \\
           &  16K &	11.5 & 4.13 & 7.2 &  12.62\\
	& 64K  & 3.9 & 42.07 &	1.7 & 120.71\\
\hline
qwen30b & 1K & 54.9 & 0.41 &26.1 & 1.55\\
	& 4K  & 51.8 & 1.63 &	25.4  & 4.27 \\
	& 16K & 44.0 & 7.37 & 20.8 & 20.3\\
	& 64K & 24.1 & 50.64 & 10.1  & 145.83\\
\hline
qwen235b & 1K & 7.7 & 37 &  & \\
	& 4K  & 7.2 & 60.4 & \multicolumn{2}{c}{OUT OF} \\
	& 16K & 7.04 & 163 & \multicolumn{2}{c}{MEMORY}  \\
	& 64K & 4.54 & 103 &   & \\
\hline
\end{tabular}
\vspace{-0.1in}
\caption{TPS and TTFT from pipelined sharding on machines \machine{cli2} and \machine{cli1} across context sizes for VRAM budget at peak capacity (which is 16G on \machine{cli2} and 12G on \machine{cli1}).}
\label{tab:bestmetrics}
\vspace{-0.1in}
\end{table}

In Table~\ref{tab:bestmetrics}, we summarize just the best TPS and TTFT metrics, corresponding to peak VRAM usage on machines \machine{cli2} and \machine{cli1}. Our review of the full data (not shown here due to space constraints) shows for VRAM budgets up to half the respective peak capacities on \machine{cli2} and \machine{cli1}, pipelined sharding delivers more than 5 TPS for all models up to 16K contexts,
enabling significant (often 3-4$\times$ and up to 13$\times$) TPS speedups across models and context sizes over the corresponding baselines. For the 64K context, at an 8G VRAM budget, only \model{qwen30b} achieves $>$5 TPS (19 TPS on \machine{cli2} and 8.75 TPS on \machine{cli1}).

\begin{figure}[htb]
\includegraphics[width=0.97\linewidth]{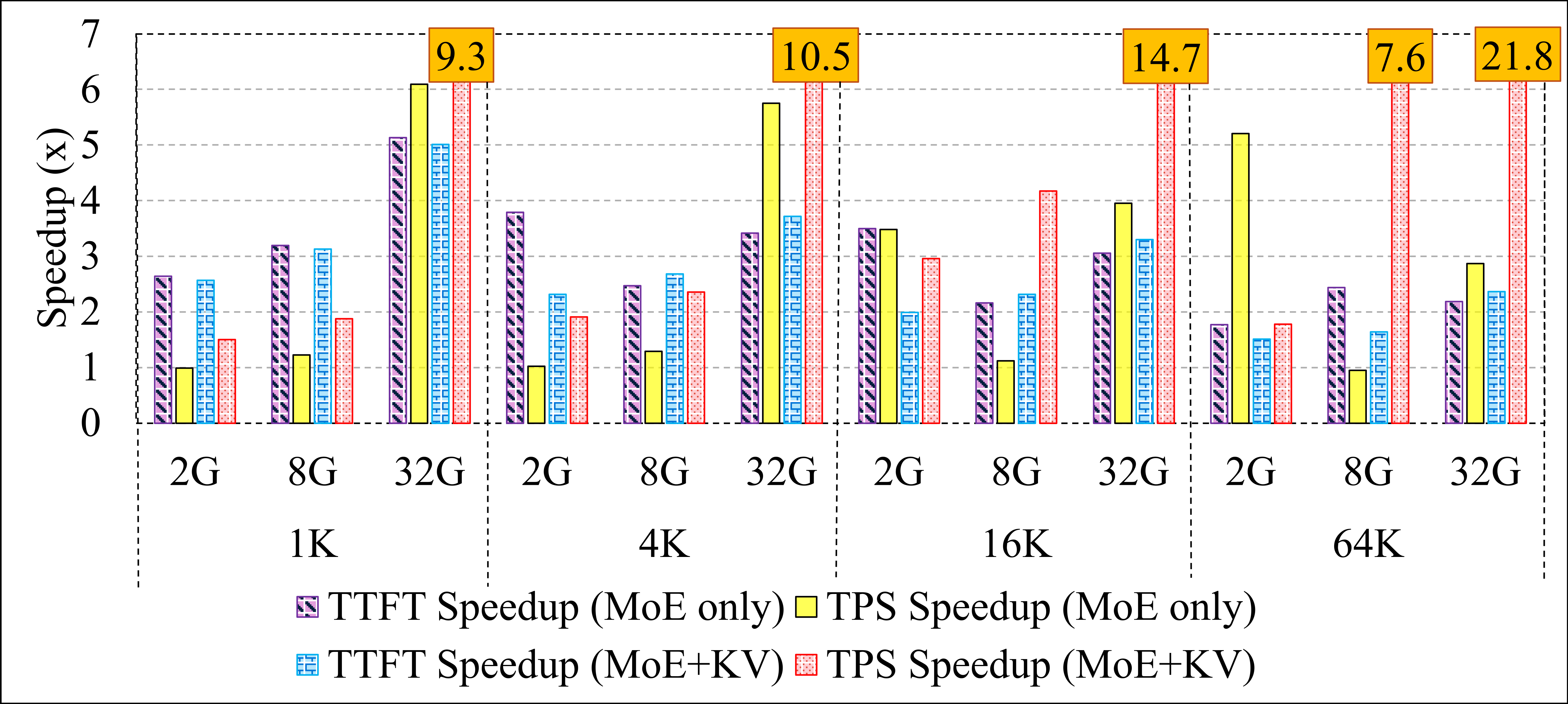}
\centering
\scriptsize
\vspace{-0.1in}
\caption{TTFT and TPS speedups from pipelined sharding vs llama.cpp's manual CPU offloading for \model{qwen30b} on \machine{cli3} across context sizes and VRAM budgets. Values $>$7$\times$ are labeled.}
\label{fig:manualoffload}
\vspace{-0.1in}
\end{figure}

{\bf Comparison with manual offloading:} In Figure~\ref{fig:manualoffload}, we compare against llama.cpp's manual CPU offloading options (MoE FFNs -cmoe and/or KV cache -kvo) for \model{qwen30b} on \machine{cli3} with 16 CPU threads. These knobs are applied atop \emph{llama-cpp-baseline} to get the corresponding baselines. 

Pipelined sharding outperforms manual MoE FFN offloading across most conditions. Manual offloading can match our automatic approach's TPS at low VRAM budgets and small contexts, but consistently degrades TTFT, demonstrating the importance of token tier based scheduling, which elegantly adapts to the high token count in an LLM's context phase by preferring the \plangpuonly{} plan.  
In the decode phase (yellow bar), as context size increases from 1K to 64K, the corresponding KV cache grows as well and competes for space in the available VRAM budget, which in turn limits the number of VRAM-mapped attention sub-layers in this baseline. However, pipelined sharding's \plandynamic{} plan maps more attention blocks to the GPU by oversubscribing the VRAM scratch area, which explains the speedups at the 2G budget for higher context sizes.  At 8G, the baseline with MoE offloading can fit all attention sub-layers and KV cache (the most latency-sensitive components) in VRAM, leaving little room for decode-phase TPS improvement.  At 32G, even though the baseline fits all the attention sub-layers and KV cache in VRAM,  forcing MoE FFNs to the CPU prevents it from using the additional VRAM for compute, whereas pipelined sharding's automatic scheduler has no such constraint and assigns MoE sub-layers to the GPU when VRAM allows. 

While offloading both MoE FFNs and KV cache avoids the above VRAM contention, the baseline suffers from poor performance due to all KV cache accesses going to sysRAM, even for GPU-mapped attention blocks, resulting in poorer baseline performance. At 64K context, our adaptive scheduler delivers superior performance, with TPS speedups reaching up to 21.8$\times$. The above experiments illustrate how our automatic scheduler adapts well to different scenarios, often exceeding the performance of manual knob settings.

{\bf VLM Performance:} Since llama.cpp does not support video inputs in its multimodal path as of this writing,  we restrict our VLM evaluation to image inputs.  Table~\ref{tab:vnemoperf} provides E2EL for \emph{llama-cpp-baseline}~(base in the table) and pipelined sharding plus VLMOpt~(psh+vopt in the table) for \model{vnemo4b} on machines \machine{cli2} and \machine{cli3}. Our solution improves E2EL by up to 1.78$\times$. It enables this IGI model to be functional at the 2G VRAM budget. Another noteworthy trend across VRAM budgets is that our solution achieves approximately the same E2EL at a VRAM budget that is 2G lower than what the baseline requires. This freed-up VRAM (at iso-performance) can be put to good use by developers to improve the performance or functionality of concurrently running applications such as video games, without regressing the latency of \model{vnemo4b}.

\begin{table}[htb]
\vspace{-0.1in}
\centering
\scriptsize
\begin{tabular}{r|rrr|rrr}
\hline
\multicolumn{1}{c}{VRAM} & \multicolumn{3}{c}{\machine{cli2}} & \multicolumn{3}{c}{\machine{cli3}} \\
\multicolumn{1}{c}{Budget} & base & psh+vopt & speedup & base & psh+vopt & speedup \\
 & (s)    &   (s)       &     (x)       &  (s)    &   (s)       &     (x)   \\
\hline
2G & \emph{OOM} & 15.91 & $-$ &     \emph{OOM} &	7.32	& $-$ \\
4G &	16.47	& 11.30 & 1.46  & 7.61	& 5.78 & 1.32	\\
6G &	11.36	& 6.37 & 1.78   & 5.64	& 3.24 & 1.74	 \\
8G & 7.56 & 4.47	& 1.69   & 3.74	& 2.10 & 1.78	\\
10G & 1.22	& 1.11 &	1.10   & 1.21 & 0.87 & 1.39	\\
\hline
\end{tabular}
\vspace{-0.1in}
\caption{E2EL performance with pipelined sharding + VLMOpt for \model{vnemo4b} on \machine{cli2} and \machine{cli3}.}
\vspace{-0.1in}
\label{tab:vnemoperf}
\end{table}

\begin{table}[htb]
\centering
\scriptsize
\begin{tabular}{c|rrr|r|rrr|r}
& \multicolumn{4}{c}{machine \machine{cli2}} & \multicolumn{4}{c}{machine \machine{cli3}}\\
\hline
image & \multicolumn{3}{c}{pshard+vlmopt} & vLLM &  \multicolumn{3}{c}{pshard+vlmopt} & vLLM  \\
res & 2G & 8G  & 14.5G  & max & 2G & 8G	& 14.5G & 20G  \\
\hline
480p	& 33.7  & 17.6  & 2.3 & \emph{OOM} & 13.6 & 8.2 & 1.8 & 8.7 \\
720p	& 35.2 & 18.2 & 2.9 & \emph{OOM} & 13.8 & 8.5 & 2.0 & 8.7  \\
1080p & 34.9 & 19.2  & 3.7 & \emph{OOM} & 14.7 & 9.1 & 2.6 & 9.0  \\
1440p & 40.4  & 21.6 & 5.9 & \emph{OOM} & 18.7 & 10.6 & 4.1 & 9.5  \\
\hline
\end{tabular}
\vspace{-0.1in}
\caption{E2EL (\emph{s}) with pipelined sharding + VLMOpt  for \model{cr1} across image resolutions and VRAM budgets  on \machine{cli2} and \machine{cli3}.}
\label{tab:cr1e2e}
\end{table}

Table~\ref{tab:cr1e2e} presents E2EL for \model{cr1}. First, the table shows that we have accomplished our primary goal of enabling physical AI developers and users to run \model{cr1} at minuscule VRAM budgets on client systems. Specifically, pipelined sharding and VLMOpt (pshard+vlmopt) together reduce VRAM demand of \model{cr1} by up to 10$\times$ (20GB to 2GB), with efficient performance at any given VRAM budget. For pipelined sharding, the last column ends at VRAM budget of 14.5G as that is the peak demand for \model{cr1} for our solution.

\begin{table}[htb]
\centering
\scriptsize
\begin{tabular}{c|rrr|rrr}
\multicolumn{1}{c}{Image Res} &  \multicolumn{3}{c}{\machine{cli2}: VRAM Budgets} & \multicolumn{3}{c}{\machine{cli3}: VRAM Budgets} \\
\hline
 & 4G	 & 8G & 14.5G  & 4G & 8G & 14.5G\\
\hline
480p  & 1.3 & 1.3 & 3.5 &  1.2  & 1.3 &  2.5 \\
720p & \emph{OOM} & 1.6 & 4.6 & \emph{OOM} & 1.5 & 3.2  \\
1080p & \emph{OOM} & \emph{OOM} & 9.0 & \emph{OOM} & \emph{OOM} & 6.7  \\
1440p & \emph{OOM} & \emph{OOM} & \emph{OOM} & \emph{OOM} & \emph{OOM} & \emph{OOM} \\
\hline
\end{tabular}
\vspace{-0.1in}
\caption{E2EL Speedups ($\times$) for \model{cr1} on \machine{cli2} and \machine{cli3} with pipelined sharding + vlmopt. Here, OOM indicates that our baseline ran out of memory and so reporting a speedup is not possible.}
\label{tab:cr1e2elspeedups}
\end{table}

Though, on the face of it,  our solution appears to perform better than vLLM at peak VRAM, our investigation shows this has more to do with vLLM's handling of multimodal inputs than our optimization.
Since vLLM does not run at VRAM budgets below 20GB and given the above performance anomaly, we can only compare our solution's performance against \emph{llama-cpp-baseline} at various VRAM budgets. However, this baseline  too fails to handle \model{cr1} for various input resolution-VRAM budget combinations (specifically, 1440p does not work at any VRAM budget, 1080p does not work at 10G or less, 720p at 4G or less, and 480p at 2G).  Table~\ref{tab:cr1e2elspeedups} summarizes a few E2EL speedups on machines \machine{cli2} and \machine{cli3}. 

From the above experiments, we can see that pipelined sharding and VLMOpt enable efficient inference for diverse models across VRAM budgets, delivering good TTFT, TPS, and E2EL speedups.

{\bf Profiler effectiveness:} We ran an oracle comparison across 105 configurations spanning 2 models (\model{nemo8b}, \model{qwen30b}), 2 PCIe generations (gen3, gen5), 2 thread counts (1, 16), 2 context sizes (4K, 16K), and multiple VRAM budgets on \machine{cli3}. For each configuration, we measured the actual TPS of all three strategies (\plangpuonly, \planstatic, \plandynamic) and compared against our planner's selection. The planner selects the optimal strategy 100\% of the time (105/105 configurations). While individual schedule latency predictions show median error around 10\%, this does not affect selection correctness because relative ordering between strategies is accurately predicted.
Across 105 configurations, no single strategy was universally optimal: \planstatic\ won in 76 cases (72\%), \plandynamic\ in 19 cases (18\%), and \plangpuonly\ in 10 cases (10\%). For instance, \plangpuonly\ outperforms alternatives under constrained CPU threads (1 thread, PCIe Gen5), while \planstatic\ dominates when CPU threads are plentiful. This validates the need for adaptive selection  and that all three strategies contribute to our overall gains.

\begin{figure}[htb]
\centering
\scriptsize
\includegraphics[width=\linewidth]{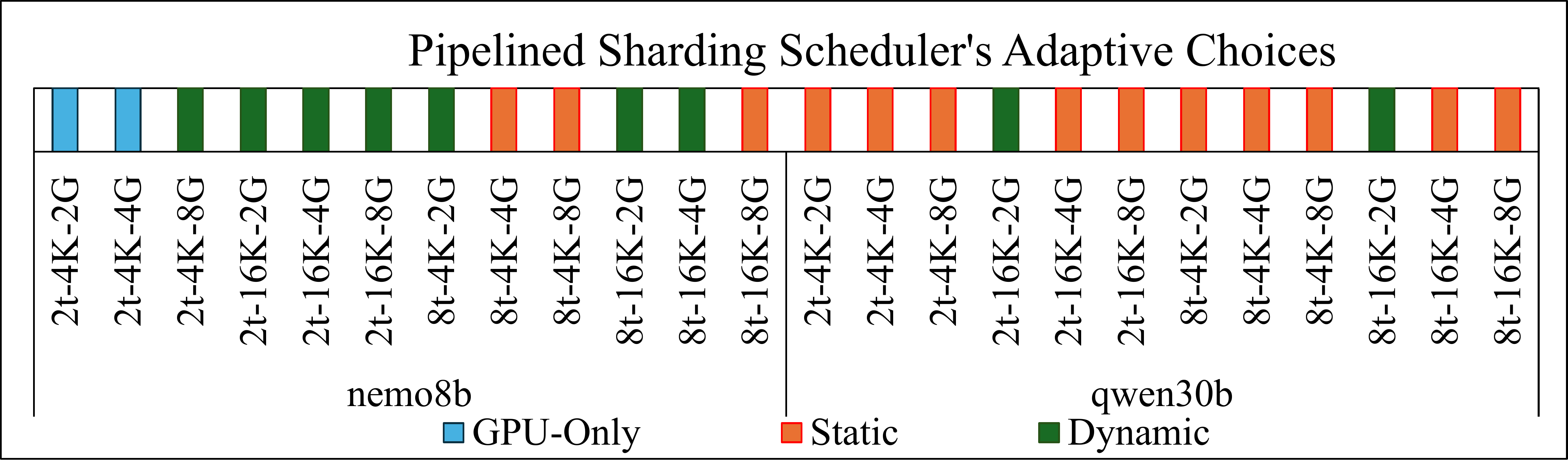}
\vspace{-0.2in}
\caption{Schedule choices adapting to LLM and system conditions.}
\label{fig:profiledrivensched}
\vspace{-0.1in}
\end{figure}

Figure~\ref{fig:profiledrivensched} shows the scheduling decisions for \model{nemo8b} and \model{qwen30b} across CPU thread counts (2t and 8t), context sizes (4K and 16K), and VRAM budgets (2G, 4G, and 8G). With fewer threads (2t), the scheduler predominantly selects \plangpuonly\ execution. With more threads (8t), it shifts to \planstatic\ or \plandynamic\ plans that leverage CPU resources, demonstrating adaptivity to system conditions.

\begin{figure}[htb]
\centering
\scriptsize
\includegraphics[width=\linewidth]{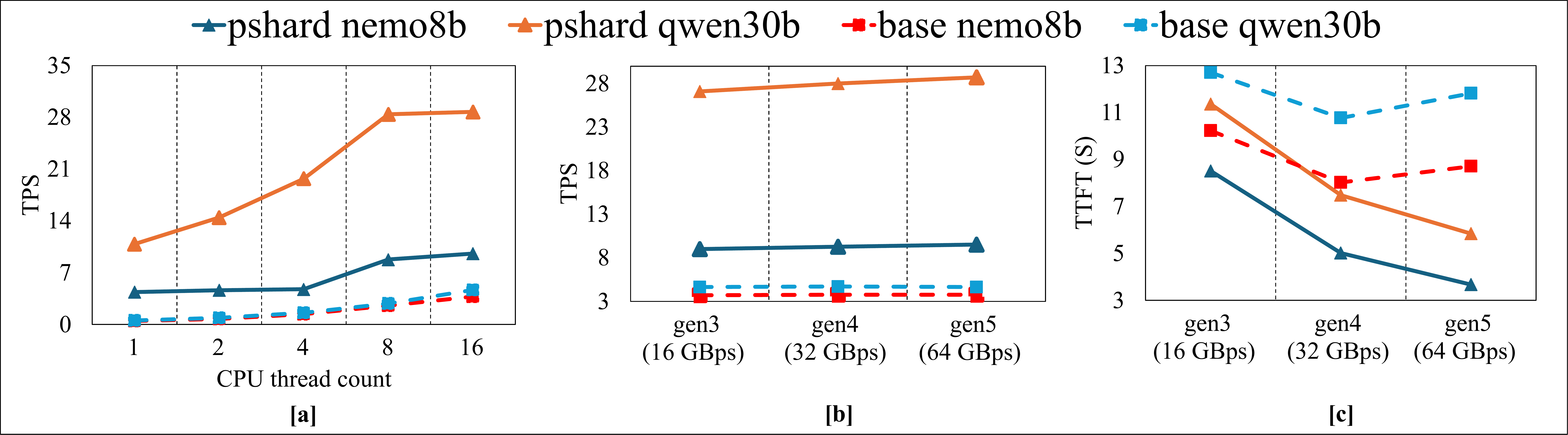}
\vspace{-0.2in}
\caption{Sensitivity studies on \machine{cli3}. (a) TPS vs CPU thread count  at 8G VRAM and 16K context. (b) TPS and (c) TTFT vs. PCIe generation at 16K context and 8G VRAM with 16 CPU threads. Lower TTFT is better.}
\label{fig:schmoos}
\vspace{-0.1in}
\end{figure}

{\bf Sensitivity studies:} Figure~\ref{fig:schmoos}a shows baseline and pipelined sharding TPS across CPU thread counts on \machine{cli3} at 8G VRAM and 16K context. The performance gap widens with thread count, growing from 10.3 TPS at 1 thread to 24.0 TPS at 16 threads for \machine{qwen30b}, demonstrating effective CPU thread utilization. This allows our framework to execute alongside other CPU applications while maintaining performance advantages.

Figures~\ref{fig:schmoos}b and ~\ref{fig:schmoos}c show TPS and TTFT when throttling PCIe bandwidth on \machine{cli3} via system BIOS settings, from gen5 (64 GBps) to gen3 (16 GBps) at 16K context and 8G VRAM. Pipelined sharding's TTFT speedup nearly doubles from gen3 to gen5 (1.2$\times$ to 2.4$\times$) due to faster weight streaming. Baseline shows moderate TTFT improvement (13-17\%) as intermediate outputs transfer over PCIe, but minimal TPS variation. Pipelined sharding's TPS also shows modest gains (4-6\%) as the scheduler selected a Static plan, entailing minimal PCIe activity, during decode for this inference setting across PCIe generations. Across both experiments, trends indicate that performance uplift scales with increased resources, which augurs well for pipelined sharding on more powerful future client systems. 

\begin{figure}[htb]
\centering
\scriptsize
\vspace{-0.1in}
\includegraphics[width=\linewidth]{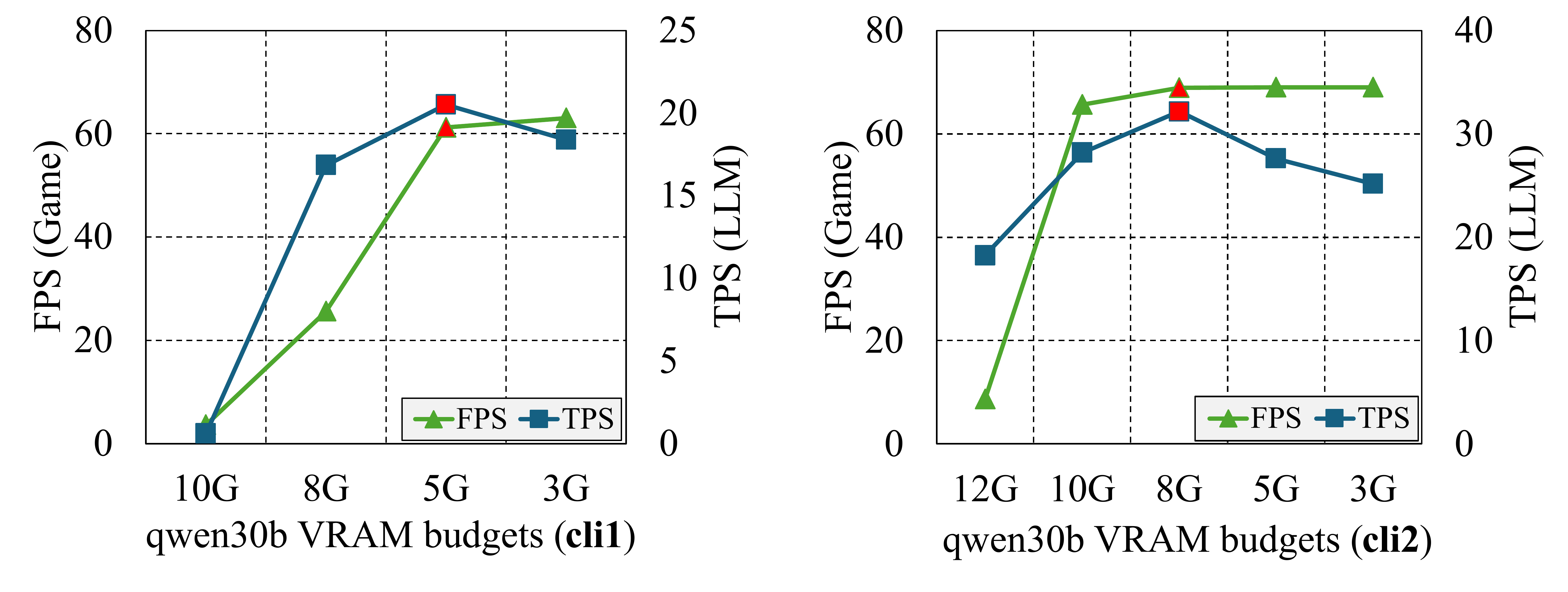}
\vspace{-0.3in}
\caption{
\model{qwen30b} TPS and Cyberpunk 2077 FPS during concurrent execution on \machine{cli1} (1080p) and \machine{cli2} (4K) across \model{qwen30b} VRAM budgets. Red markers highlight pareto-optimal budgets.}
\label{fig:vidgame}
\vspace{-0.05in}
\end{figure}

{\bf With a concurrent video game:} In Figure~\ref{fig:vidgame}, we show the impact on TPS and frames per second (FPS) for \model{qwen30b} running concurrently alongside the DX12 video game, Cyberpunk2077 with mainstream settings on \machine{cli1} (1080p) and \machine{cli2} (4K).  The x-axes show various VRAM budgets allocated to \model{qwen30b}. At high VRAM budgets, the concurrently running game cannot fit all its assets in VRAM and has to spill to sysRAM. This causes game frame time to increase significantly and due to poor preemption behavior of slow-running graphics calls, the LLM does not get enough GPU compute cycles, leading to low TPS. 
Constraining the LLM to smaller budgets with pipelined sharding allows the game to keep more (and eventually all) of its assets in VRAM.
Improved game performance leads to faster preemption, allowing timely LLM inference.  Both TPS and FPS are high at a pareto-optimal VRAM budget sweet spot, underscoring the importance of our solution.

Here, we ran the LLM and the game as separate processes. They share available VRAM but do not share GPU compute due to time-sliced execution (no concurrent kernel execution). We expect future applications to also use time-slicing to guarantee quality of service (QoS) for individual tasks. Finally, while we identified the sweet spots above through manual experiments, perfect automatic co-scheduling of two demanding GPU processes will require a dedicated scheduler such as NVIDIA's AI Management Processor~\cite{blackwellwhitepaper} or invoking the \xlm\ from the game's process and handling with CUDA-in-Graphics~(CiG). Studying these possibilities is outside the scope of this work.

{\bf Batching:} While all the experiments thus far used a batch size of 1, we next evaluate pipelined sharding for multi-request batches. 
We evaluate both dense~(\model{nemo8b}) and MoE~(\model{qwen30b}) models  with 1K and 4K per-request context sizes at VRAM budgets of 4G, 8G, and 16G on \machine{cli3}. 
llama.cpp supports both unified~(\ukv) and non-unified~(\nukv) KV caches for multi-request batches. We evaluate both options and present pipelined sharding's raw TPS with \nukv\ as well as additive deltas to get TPS with \ukv\ in  Table~\ref{tab:batchtps}.  

\begin{table}[t]
\centering
\scriptsize
\begin{tabular}{c|r|rrrr}
\hline
Model & VRAM & \multicolumn{1}{c}{bs=1} & \multicolumn{1}{c}{bs=4} & \multicolumn{1}{c}{bs=16} & \multicolumn{1}{c}{bs=64} \\
(Context) & Budget & nukv / $\Delta$ & nukv / $\Delta$ & nukv / $\Delta$ & nukv / $\Delta$\\
\hline
\hline
\model{nemo8b} & 4G & 9 /  0 & 32 /  0 &52 /  1&104 /  5 \\ 
(1K) & 8G & 12 /  1 & 44 /  1 &73 /  2&122 /  6\\ 
& 16G& 71 / -1 & 198 / 14 &205 / 25&202 / 11\\ 
\hline
\model{nemo8b} & 4G & 8 /  0 & 27 /  0 &31 /  0 &39 / -4\\ 
(4K) & 8G & 12 /  0 & 37 /  0 &38 /  0&42 / -4 \\ 
& 16G & 64 /  0 & 105 /  2 &73 /  2&50 / -7 \\ 
\hline
\model{qwen30b} & 4G & 27 /  0 & 71 /  5 &122 / 12&101 / 13 \\
(1K) & 8G & 31 /  0 & 76 /  6 &144 / 18&169 /  7 \\
& 16G & 47 / -5 & 96 / -2 &229 / 30&270 / 19 \\
\hline
\model{qwen30b} & 4G & 27 /  0 & 70 /  3 &30 /  0&56 /-13 \\
(4K) & 8G & 30 /  0 & 75 /  4 &119 /  4&50 / -7 \\
& 16G & 41 / -1 & 88 /  5 &163 / 12&62 /-11 \\
\hline
\end{tabular}
\vspace{-0.1in}
\caption{TPS for pipelined sharding for various batch sizes on \machine{cli3} across VRAM budgets and context sizes. Data cells represent "non-unified KV TPS / additive $\Delta$ to get unified KV TPS". }
\label{tab:batchtps}
\vspace{-0.1in}
\end{table}

\begin{figure}[h]
\vspace{-0.1in}
\centering
\scriptsize
\includegraphics[width=\linewidth]{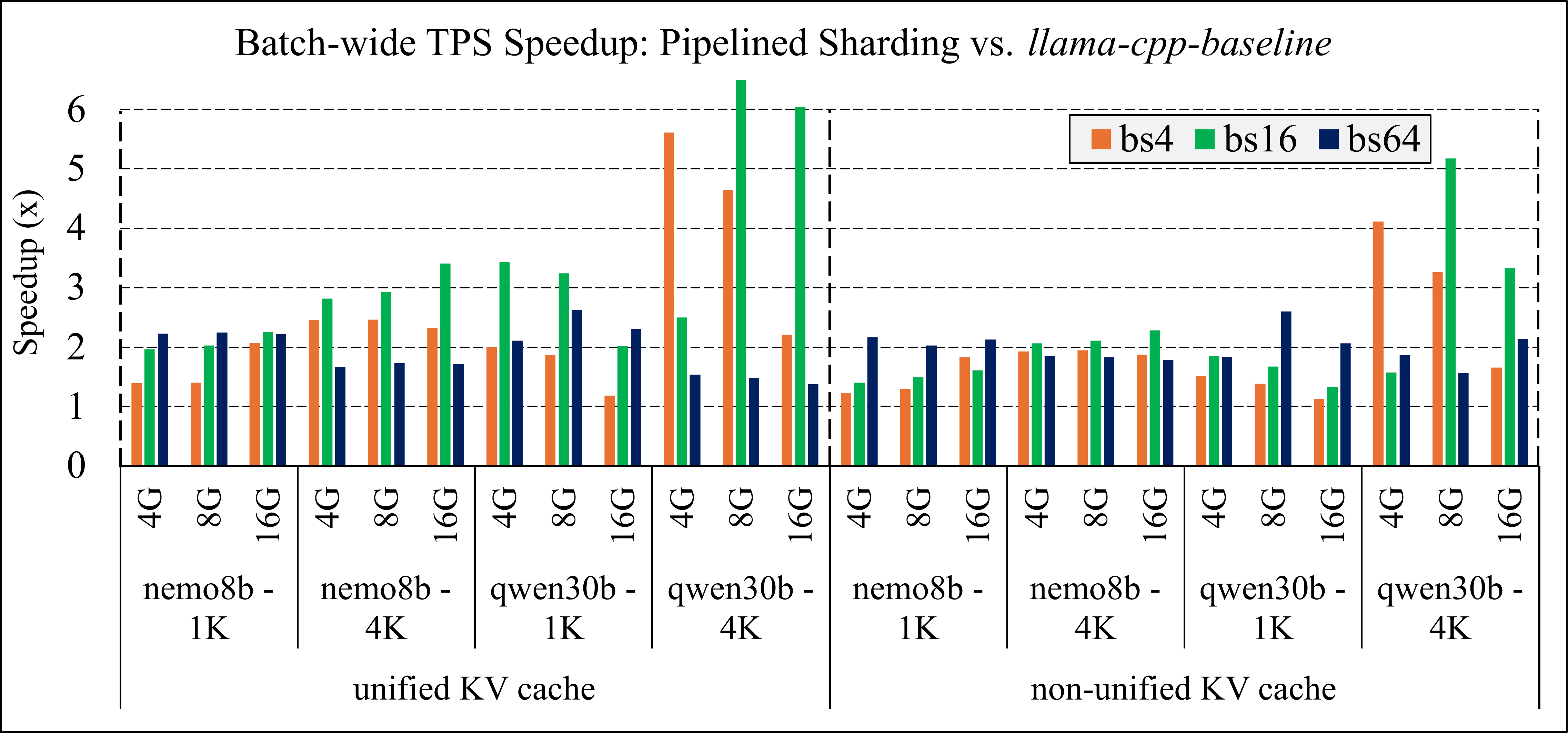}
\vspace{-0.2in}
\caption{Pipelined sharding batched mode performance on \machine{cli3} across batch sizes (4, 16, 64), context sizes (1K and 4K), and VRAM budgets (4G, 8G, 16G).  Batch-wide TPS improves by \batchavg\ on average and up to \batchmax{}.}
\label{fig:batching}
\end{figure}

With 1K context, TPS generally scales with batch size up to 64, with \model{qwen30b} reaching 289 TPS (\ukv). 
For 4K, for \model{qwen30b} at all budgets and \model{nemo8b} at 16G, TPS drops with larger batch sizes at the point where the aggregate KV cache overshoots the VRAM budget, leading to reduced KV pinning. For \model{nemo8b} at 4G and 8G, since KV cache overshoots VRAM budget at all batch sizes, TPS scales modestly with batch size.
Figure~\ref{fig:batching} shows that pipelined sharding delivers good batch-wide TPS speedups across batch sizes with both \ukv\ and \nukv, relative to their respective \emph{llama-cpp-baseline} runs. 
Across models and settings (including batch sizes), on average, we see a \batchavg\ speedup.  At 4K context, \model{qwen30b} with batch size 16 speeds up by \batchmax\ and 5.2$\times$ for \ukv\ and \nukv, respectively. 
Speedups tend to increase with batch size because the baseline's CPU-executed tensors grow with batch size, slowing its CPU-side performance relative to pipelined sharding's token-tier selection that adapts by shifting work to the GPU with \plangpuonly\ or \plandynamic{} plans. 
These results demonstrate that pipelined sharding yields strong performance not only in interactive mode, but also in batched mode.

%% file: discuss.tex
\section{Lessons Learned}
\label{sec:discuss}

We identify seven main vendor-agnostic takeaways from our work that practitioners may find useful. 

{\bf Lossless inference:} Early feedback from client AI developers on the importance of lossless inference of high-accuracy models across SKUs, steered us away from lossy approaches and toward pipelined sharding. 

{\bf Contention-aware profiling:} 
We found that directed test profiles (e.g., PCIe-only or CPU-only) are poor proxies for hybrid schedule performance, since they miss contention between PCIe and CPU for shared resources like the memory controller. We overcame this by profiling under concurrent load, which produced contention-adjusted throughput metrics that improved cost estimation accuracy.

{\bf Token-tier-based generalized scheduler:} On heterogeneous hardware, the optimal distribution of inference work shifts with the number of tokens processed per iteration. At high token counts, compute dominates and data movement cost is amortized, so offloading all work to the accelerator is optimal. At low token counts, compute shrinks and data movement becomes the bottleneck, so distributing work across resources to reduce transfers and overlap the remaining ones delivers better throughput. Organizing schedules into token tiers and selecting an appropriate schedule per inference iteration naturally captures all operating modes, from single-request interactive use to multi-request batched scenarios. It was easier to design a generalized scheduler than it was to optimize for interactive or multi-request modes individually. 

{\bf Homogeneous scheduling units:} As scheduling units, we use sub-layers formed by splitting LLM layers at the attention-FFN boundary because arithmetic intensity changes there. Since such scheduling units are computationally homogeneous, prioritized backend assignment based on their arithmetic intensities (guided by roofline models) was effective for both dense and MoE models. 

{\bf Decoupled profiling, planning, and inference:} Separating profiling (install time), planning (once per model invocation), and inference into distinct phases eliminated overhead from the critical inference path. 

{\bf Simplified scheduling:} In a production setting, evaluating every possible sub-layer-to-hardware mapping is expensive. Instead, we first generate a small set of schedules from first principles, each representing a fundamentally different strategy to map work to available resources. We then select from among them based on benchmark-profile-driven cost estimates. This simplified approach led to fast, robust, and accurate scheduling decisions. 

{\bf Extensibility:} We designed for extensibility. As new models introduce operations with different arithmetic intensities or as hardware capabilities evolve (e.g. NPUs in modern CPUs), 
adding new kernels to the benchmark suite or new schedule strategies to the planning phase allows cost estimation and schedule selection to adapt automatically.
Similarly, more schedules can be included for ``estimate and pick'' in the planning phase, potentially enabling better inference performance at a modest planning cost that is paid prior to the critical inference loop.

%% file: conc.tex
\section{Conclusion}
\label{sec:conc}

Pipelined sharding and VLMOpt make efficient, VRAM-constrained \xlm\ inference accessible to client AI developers and users. 
Pipelined sharding can automatically adapt to system configurations, inference phases, context sizes, batch sizes, and VRAM budgets, thanks to its benchmark-profile-driven schedule cost estimation and token tier design. 
Our contributions not only help improve the performance and VRAM consumption of current IGI SDK models, but also  enable larger, higher-accuracy models to run efficiently in user-specified VRAM budgets, while achieving TTFT and TPS metrics that are appropriate for in-game interactive use cases. Our work paves the way for game developers to use these more accurate models immediately via IGI SDK's plugin feature, but also potentially opens the door to including such models as part of the core IGI models in future releases of the SDK.
Our work reduces the NVIDIA Cosmos-Reason1 VLM's VRAM demand by 10$\times$ compared to its current vLLM baseline, enabling high-resolution CR1 inference on clients, while also improving its performance across various VRAM budgets, which we hope will democratize physical AI development and deployment.

%% file: artifactappendix.tex
\section{Artifact Appendix}
\label{sec:artifactappendix}
\subsection{Abstract}

Our paper presents two optimizations - pipelined sharding and VLMOpt - to achieve efficient \xlm\ inference at user-specified VRAM budgets on client systems. We have implemented them on top of llama.cpp tag  \href{https://github.com/ggml-org/llama.cpp/tree/b6097}{b6097}. This artifact contains the source code, build scripts, model download scripts, and scripts to reproduce key results, namely, Table~\ref{tab:tps}~(TPS and TTFT across VRAM budgets),  Figure~\ref{fig:bestresults}~(TTFT/TPS/E2EL speedups over \emph{llama-cpp-baseline}), Table~\ref{tab:cr1e2elspeedups}~(E2EL speedups for Cosmos-Reason1 with VLMOpt), Table~\ref{tab:batchtps}~(TPS vs batch size across VRAM budgets), and Figure~\ref{fig:batching}~(pipelined sharding batch scaling for \model{qwen30b}). Automated profiling, benchmarking, and result-parsing scripts are included for each experiment.

We recommend testing on an x64 system with one of NVIDIA RTX 5090/5080/4090/4080 or A100 GPUs. 
We expect speedup trends across these experiments to be within a minimal tolerance of the reported speedups. Our validation scripts will print PASS (or not) along with error margins. 

\subsection{Artifact Check-list (Meta-information)}
\begin{itemize}[label={}, leftmargin=0pt, itemindent=0pt, labelsep=0pt]
\item {\bf Program:} C++ implementation of Pipelined Sharding (contention-aware profiling, priority-based tensor placement, phase-adaptive scheduling) and VLMOpt in llama.cpp. PowerShell and Bash reproduction scripts to reproduce all the aforementioned results.
\item {\bf Compilation:} CMake with CUDA support. On Windows, CMake also generates a Visual Studio solution (`.sln`) for building via Visual Studio 2022+. Build instructions are provided in the README.
\item {\bf Binary:} Pre-built Windows and Ubuntu Linux x64 release binaries available at \url{https://github.com/deepshnv/pipeshard-mlsys26-ae/releases}.
\item {\bf Model:} Six GGUF models (see Table 2 in the paper). Download scripts provided (`download\_models.ps1'/ `download\_models.sh'). Models are hosted on Hugging Face and NVIDIA Developer.
\item {\bf Data set:} Context prompt files (1K–64K tokens) and a test image are included in the repository under `paper\_results/'.
\item {\bf Run-time environment:} Windows or Linux with a discrete GPU. NVIDIA GPU with CUDA Toolkit 12.8+ is recommended but not strictly required; llama.cpp supports multiple GPU backends (CUDA, Vulkan, HIP, Metal, SYCL).
\item {\bf Hardware:} We recommend testing on an x64 system with one of NVIDIA RTX 5090/5080/4090/4080 or A100 GPUs.
\item {\bf Execution:} All benchmarks execute on the GPU and the host CPU. Hardware profilers run first to characterize the system, then inference benchmarks run the models.
\item {\bf Metrics:} Time-to-first-token (TTFT), tokens per second (TPS), end-to-end latency (E2EL), image encode/decode time (for VLMs), peak VRAM usage, and speedups relative to baselines.
\item {\bf Output:} CSV files with per-configuration metrics and speedups, printed summary tables on the console.
\item {\bf Experiments:} Reproduction of Table 4, Table 5, Table 8, Table 9, and Figure 2 from the paper.
\item {\bf How much disk space required (approximately)?} \textasciitilde175 GB for all model weights. \textasciitilde3.5 GB for source code and build artifacts. \textasciitilde6 GB for CUDA SDK installation.
\item {\bf How much time is needed to prepare workflow (approximately)?} \textasciitilde20 minutes to build from source; ~1$-$3 hours to download all models depending on internet speed.
\item {\bf How much time is needed to complete experiments (approximately)?} Table 4: \textasciitilde1$-$2 hours (112 runs). Table 5: \textasciitilde30 minutes (16 runs). Table 8: \textasciitilde1 hour (16 runs). Table 9: 1 hour (12 runs). Figure 2: \textasciitilde6$-$8 hours (up to 256 paired runs). Total: \textasciitilde9$-$13 hours for the full suite. These estimates are based on an RTX 5090 GPU and will vary with hardware; slower GPUs or CPUs may take proportionally longer. Individual tables can be run independently.
\item {\bf Publicly available?} Yes.
\item {\bf Code licenses (if publicly available)?} MIT License.
\item {\bf Data licenses (if publicly available)?} Model weights are subject to their respective licenses (Apache 2.0 for Qwen, NVIDIA License for minitron/CR1).
\item {\bf Archived (provide DOI)}? \href{https://zenodo.org/records/19436383}{doi url}
\end{itemize}
\subsection{Description}
\subsubsection{How Delivered}

The artifact is hosted on GitHub at \url{https://github.com/deepshnv/pipeshard-mlsys26-ae} and archived on Zenodo. The repository contains the full source code (a fork of llama.cpp), build instructions, model download scripts, reproduction scripts, and reference result images from the paper.

\subsubsection{Hardware Dependencies}

An x86\_64 machine with NVIDIA RTX 5090 GPU (ideally) or any RTX 40/50 series GPU or an NVIDIA compute node (A100 or newer). The reproduction scripts work on any VRAM size-- the paper's VRAM-budget sweep (2G-32G) is parameterized and reviewers can adjust the range to match their hardware.

\subsubsection{Software Dependencies}
\begin{itemize}[label={}, leftmargin=0pt, itemindent=0pt, labelsep=0pt]
\item {\bf GPU driver appropriate for your hardware (recommended:} Game Ready / Studio Driver for NVIDIA RTX 40/50 series GPUs and the R570-Server driver for NVIDIA compute GPUs (A100 or newer).
\item {\bf GPU compute toolkit (recommended:} CUDA Toolkit 12.8+ for NVIDIA RTX GPUs);
\item {\bf Others:} CMake 3.20+, Visual Studio 2022+ (Windows) or GCC/Clang (Linux), Python 3.12+ with `huggingface\_hub[cli]' (for model downloads)
\end{itemize}

\subsection{Installation}

{\bf Option A: Build from Source}

\begin{lstlisting}[caption={}]
> git clone https://github.com/deepshnv/pipeshard-mlsys26-ae.git
> cd pipeshard-mlsys26-ae 
> cmake -B build -DGGML_CUDA=ON -DLLAMA_CURL=OFF
> cmake --build build --config Release -j16
\end{lstlisting}

{\bf Option B: Pre-built Binaries (Windows x64 or ubuntu 24.04 x64)}

Download the release archive from \url{https://github.com/deepshnv/pipeshard-mlsys26-ae/releases} and extract to `build/bin/Release/'.

\subsection{Download Models}

\begin{lstlisting}[caption={}]
# For Windows, from powershell:
> .\download_models.ps1

# For Linux/macOS:
$ ./download_models.sh
\end{lstlisting}

\subsection{Experiment Workflow}

The exhaustive list of steps (with various options) to reproduce our results are available in our \href{https://github.com/deepshnv/pipeshard-mlsys26-ae?tab=readme-ov-file\#reproducing-mlsys26-paper-results}{README}.
We will summarize that briefly here. All scripts to reproduce our results are located in `paper\_results/'. 
We have also provided a convenient top level script that can be invoked as follows to run all experiments with one simple command.
\begin{lstlisting}[caption={}]
> .\run_all_repro.ps1 # Windows
$ ./run_all_repro.sh   # Linux
\end{lstlisting}

Optionally, an evaluator can also invoke per-experiment scripts (repro*.ps1 and repro*.sh for Windows and Linux, respectively). Each of these scripts runs our benchmark performance profiler (unless `--skip-profiling' is specified), launches inference runs, parses metrics from the output, and writes results to a CSV file. The top-level file simply invokes these repro* scripts in order. After you run the top-level script, you should expect to find per-experiment outputs as follows.

All outputs will be found in the {\small \tt paper\_results} directory. In that directory, the evaluator can find the results for TPS and TTFT across VRAM budgets (as in Table~\ref{tab:tps}) in {\small \tt table4\_results.csv},
E2EL speedups for Cosmos-Reason1 VLM (as in Table~\ref{tab:cr1e2elspeedups})  in {\small \tt table8\_results.csv}, 
and TTFT/TPS/E2EL speedups from pipelined sharding (as in Figure~\ref{fig:bestresults}) in {\small \tt figure2\_results.csv}.
Results for raw batching TPS (as in Table~\ref{tab:batchtps}) and batch scaling speedups (as in Figure~\ref{fig:batching}) can be found in {\small \tt table9\_results.csv} and  {\small \tt figure7\_results.csv}, respectively. 

\subsection{Evaluation and Expected Results}

While absolute performance numbers will vary across hardware; we expect relative speedup trends to remain consistent. 
The following command line can be used to verify the speedups achieved on the test machine against the values reported in the paper.
\begin{lstlisting}[caption={}]
> python3 compare_all_results.py
\end{lstlisting}
This script prints PASS for each speedup within 90\% of the paper's value, or the actual achieved ratio (e.g., 0.76x of paper) for those outside that threshold. Comparisons are skipped for experiments for which CSVs have not been generated yet.

{\bf Table~\ref{tab:tps}:} TPS should increase monotonically with VRAM budget for a given model and context size. At 2G VRAM, even \model{qwen235b} (77 GB on disk) should achieve interactive TPS ($>$5) for contexts up to 16K.

{\bf Figure~\ref{fig:bestresults}:} TTFT speedups are expected to be in [1.5$\times$, 6.7$\times$]. TPS speedups can go up to \textasciitilde30$\times$ for \model{qwen235b} at 64K context. E2EL speedups should range from approximately 1.2$\times$ to \textasciitilde4.3$\times$. These trends should be consistent across hardware even though absolute values may differ.

{\bf Table~\ref{tab:cr1e2elspeedups}:} E2EL speedups for VLMOpt over the baseline should be $>$1$\times$ for configurations where the baseline fits in VRAM. Higher resolutions at lower VRAM budgets show OOM for the baseline, demonstrating VLMOpt's ability to run previously unrunnable configurations.

{\bf Figure~\ref{fig:batching}:} Speedups from batched processing of requests are expected to be between [1$\times$, 2$\times$] with up to 16 requests and potentially scale to \textasciitilde3$\times$ with 64 concurrent requests at VRAM budgets of 8G and 16G.

\subsection{Methodology}
\begin{itemize}[label={}, leftmargin=0pt, itemindent=0pt, labelsep=0pt]
\item Submission \href{https://ctuning.org/ae/submission-20190109.html}{link}. Reviewing \href{https://ctuning.org/ae/reviewing-20190109.html}{link}. Artifact Review Badging \href{https://www.acm.org/publications/policies/artifact-review-badging}{link}.
\end{itemize}